# Atomistic simulations of the deformation behavior of an Nb nanowire embedded in a NiTi shape memory alloy

Jung Soo Lee [a], Won-Seok Ko [b,*], Blazej Grabowski [c]

[a] Research Institute of Basic Sciences, University of Ulsan, Ulsan 44610, Republic of Korea
[b] School of Materials Science and Engineering, University of Ulsan, Ulsan 44610, Republic of Korea
[c] Institute for Materials Science, University of Stuttgart, Pfaffenwaldring 55, 70569 Stuttgart, Germany

## Abstract

The influence of pre-strain and temperature on the superior properties exhibited by an Nb nanowire embedded in a NiTi shape memory alloy (SMA) are investigated via molecular dynamics simulations. To this end, a new Nb-Ni-Ti ternary interatomic potential based on the second nearest-neighbor modified embedded-atom method (2NN MEAM) is developed and employed. The origin of the unique phenomena of quasi-linear elasticity, slim hysteresis, and reduction in Young's modulus observed for pre-strained nanowire-SMA composites is uncovered. The results demonstrate the importance of plastic deformation in the embedded Nb nanowires and reveal how the deformation facilitates the just-mentioned, unprecedented phenomena. A simple and straightforwardly obtainable descriptor to correlate and monitor Young's modulus evolution during pre-straining is proposed. Furthermore, our simulations suggest that the desired Young's modulus can be obtained for a wide range of application temperatures through appropriate pre-straining.

*Corresponding author:
E-mail: wonsko@ulsan.ac.kr (Won-Seok Ko)

# 1. Introduction

In general, material properties are correlated among each other and these inherent relationships are difficult to break. Nevertheless, through careful alloy design and microstructure control, some of these intrinsic relations could be overcome, e.g., the inverse strength-ductility trade-off [1, 2], the temperature-ductility relationship [3], or the modulus-strength relationship [4]. An important, specific material example are freestanding nanowires that exhibit simultaneously high elastic strains as well as high yield strengths [5-11]. Attempts were made to improve the applicability of such nanowires by fabricating composites of nanowires embedded in bulk, but the intrinsic properties of freestanding nanowires could mostly not be matched by such composites [12-14]. The situation changed with an investigation by Hao *et al*. [15] who reported the development of a composite of Nb nanowires embedded in a NiTi shape memory alloy (SMA). This composite exhibited an extraordinary combination of mechanical properties, quite distinct from the properties of usual alloys. For example, Ref. [15] reported a high yield strength (~1.65 GPa), high elastic strain limit (~6%), low Young's modulus (~28 GPa), and quasi-linear elasticity after a pre-straining treatment with a tensile strain cycle of 9.5%, indicating the capabilities of this composite as a functional and biomedical material.

Studies soon followed [16-21], hunting for the origin of the extraordinary properties of SMA embedded Nb nanowires, by using in-situ high-energy X-ray diffraction [18] and in-situ transmission electron microscopy [20]. An important concept, referred to as *lattice strain matching*, was conceived and experimentally demonstrated [20, 22]. According to this concept, the lattice of the nanowire in the immediate vicinity of stress-induced martensite elastically stretches to match the lattice strain of the martensite. Although this concept could shed some light on the highly elastic nature of the nanowires embedded in a SMA matrix, the fundamental atomistic origin of the superior properties of such composites is so far not clear.

For conventional, pure SMAs (i.e., without an embedded nanowire), it is well known and understood that these materials exhibit pseudoelasticity, i.e., the ability to fully recover from large strains (exceeding by far the usual elastic limit), at appropriate temperatures. The pseudoelastic stress-strain curve of pure SMAs features a significant plateau and hysteresis (Fig. 1a), both of which originate in the special microstructural evolution during straining, i.e., in the phase transformation from austenite to martensite (Fig. 1d). For adequately pre-strained SMAs that contain embedded Nb nanowires, the stress-strain response is not pseudoelastic anymore, but *quasi-elastic* (Fig. 1b). A linear increase upon loading is followed by a non-linear behavior during unloading resulting in a hysteresis, albeit much *slimer* than for pure, pseudoelastic SMAs. A quasi-linear elastic behavior is also clearly distinct from the linear elastic response of "usual" materials (Fig. 1c), and this is not only due to the particularity of the unloading stage but also due to the much larger strain magnitudes that can be reached with quasi-linear elastic materials.

Quasi-linear elasticity is intriguing because, considering the loading curve, the material does not seem to have undergone any plastic deformation. However, depending on the exact extent of straining or pre-processing, either a complete recovery or minute permanent strain (irrecoverable strain) can be observed [15, 23-27]. In deformed SMAs, the phenomenon was theoretically explained using Landau free-energy landscapes of different martensite variants [27, 28]. The local stress fields of defects generated through the deformation, such as dislocations and lattice distortions, tilt the free-energy landscapes of the martensite variants to either suppress or stabilize certain variants. Hence, extremely small and well dispersed variants of martensite, known as nanodomains, are expected to form. Such a microstructure leads to a continuous martensitic transformation instead of a sharp martensitic transformation that is generally expected to take place. Upon loading, the nanodomains which are stabilized by both the external load and the local stress fields of the defects grow whereas those nanodomains suppressed by the external load shrink and disappear. Furthermore, the occurrence of slim hysteresis was ascribed to the restoration of the original structural states by the suppression of new martensite domains during the unloading process, since the local stress fields associated with the defects bring back those variants of martensite which have shrunk or disappeared upon application of the load. The schematic shown in Fig. 1e indicates a hypothetical microstructure evolution according to such a description. In as far this picture of evolution really applies has not been clarified to date.

Hitherto, what is known qualitatively is that some amount of pre-strain is necessary before the composite can exhibit superior properties. Quantitative dependencies have not been assessed yet. In a study by Wang *et al*. [18], even though the Nb nanowire + NiTi SMA composite was pre-strained to approximately 8%, it did not exhibit any quasi-linear elasticity in the subsequent cycle. In another study, stress-strain curves with low pre-strain (4%) and high pre-strain (9.5%) showed a clear difference, with only the high pre-strain case exhibiting quasi-linear elasticity with nearly vanishing hysteresis and low Young's modulus [23].

The available results on pre-straining and its influence on the composite properties suggest the importance of plastic deformation. For example, an analysis based on in-situ HE-XRD revealed that plastic deformation occurs in Nb nanowires subjected to high pre-strain (9.5%) [15]. Furthermore, a phase field simulation with pre-existing dislocation loops at the interface of an Nb nanowire and a NiTi matrix demonstrated the decisive role of pre-strain by revealing how dislocations impact the local stress field [23]. These results highlight the importance of pre-straining that is necessary to bring out the desired properties in composites made of nanowires embedded in a SMA matrix.

However, an influence of pre-straining was also observed for a nanowire-free SMA. A largely pre-strained (approximately 27% and greater) $Ni_{50.8}Ti_{49.2}$ SMA [23, 27] exhibited quasi-linear elasticity, small hysteresis, and low Young's modulus, showing that these properties can be achieved in the absence of a nanowire, too. The fact that also a nanowire-free SMA is able to exhibit strong changes in properties suggests that the nanowires likely play the role of accelerating and amplifying the property changes. This idea needs to be verified and quantified, i.e., how much pre-strain is required to manifestly accelerate and amplify the property changes of Nb nanowire + NiTi SMA composites, and beyond this a guideline to control the properties would be useful.

In this paper, we present molecular dynamics (MD) simulations for an Nb nanowire embedded in a NiTi SMA matrix and we investigate the influence of pre-strain and temperature via cyclic tensile loading. MD simulations are very versatile and can provide a detailed understanding of the underlying mechanisms at the microscopic scale, if a reliable interatomic potential is available. For example, the martensitic transformation of NiTi SMAs has been successfully analyzed by performing MD simulations [29-44] using a reliable interatomic potential for the binary Ni-Ti system [45]. However, since no interatomic potential for the Nb-Ni-Ti ternary system is available yet, we have developed a new potential based on the second nearest-neighbor modified embedded-atom method (2NN MEAM) [46-48]. A detailed validation is presented in the Supplementary Material and the application of this new potential to the Nb nanowire embedded in NiTi is provided in the following.

Although the previously reported properties were mostly based on single cycle pre-straining at ambient temperature, we report here on a wider range of processing conditions to offer theoretical guidance to future experiments. Especially, the role of multiple cyclic loading on the martensitic phase transformation and the property changes as a function of pre-straining are examined. We also analyze the effect of temperature because the phase transformation behavior of NiTi SMAs can change remarkably with temperature. The simulated mechanical behavior of the composite structure using the developed interatomic potential agrees well with previous experimental observations [15, 23, 27, 28]. We discuss the importance of plastic deformation and of retained martensite caused by pre-straining. A measurable descriptor to correlate and monitor Young's modulus is proposed and validated.

## 2. Methodology

To develop a ternary interatomic potential based on the 2NN MEAM formalism, the unary and binary descriptions of the constituent elements are required. The potentials of pure Ni, pure Ti, and binary Ni-Ti systems were taken from our previous work [45]. The other potentials for pure Nb, binary Nb-Ni and Nb-Ti, and ternary Nb-Ni-Ti were developed in the present work. The fitting procedure of the potential parameters is given in Appendix A. Density-functional theory (DFT) calculations were used to generate a fitting database of atomic forces and energies. The details of the DFT calculations are provided in Appendix B. The database also includes forces and energies of atomic configurations calculated with various defects and at different temperature and strain conditions. For a comprehensive benchmark of the potential against reported experimental and DFT data, we refer to the Supplementary Material.

For the MD simulations, supercells with a single Nb nanowire embedded inside of a NiTi shape memory alloy (SMA) were prepared as illustrated in Fig. 2a. First, cuboidal nanocrystalline $Ni_{50}Ti_{50}$ supercells with dimensions of ~20×17.5×17.5 nm and with three grains inside were generated employing the Voronoi construction method [49], using random positions and arbitrary crystallographic orientations for the grains, as the NiTi matrix was experimentally found to show only weak, near (111) texture [16]. Subsequently, atoms located within a cylinder of 7 nm diameter and with its axis positioned at the center of the cell parallel to the direction with the longest edge, were substituted with Nb atoms. The Nb nanowire was oriented in the well-established [110] direction along the axis of the wire [15-18, 21]. Figure 2b shows the such prepared supercell, with the colors distinguishing the three different grain orientations and the Nb nanowire. The resulting volume fraction of the wire was approximately 12.5%. The respective number of NiTi and Nb atoms was about $3.9\times10^5$ and $4.2\times10^4$.

In order to better mimic the experimentally observed conditions, i.e., that the nanowires inherently contain pre-existing dislocations, mobile dislocation loops were initially introduced into our simulated Nb nanowire. To this end, we followed the procedure proposed by Zepeda-Ruiz *et al*. [50]. Specifically, 24 vacancy loops with a diameter of 4 times the lattice constant (0.33 nm) were randomly positioned on the possible slip planes of the single crystal bcc Nb cuboid with dimension ~20×7.5×7.3 nm. After relaxation, the distributed vacancy loops typically changed into dislocation loops and their networks. Then, a cylinder with diameter of 7 nm was extracted. This cylinder was inserted into cylindrically hollow cuboidal nanocrystalline NiTi supercell with three grains. The details of cell configurations designated to study the effect of dislocation density are provided in the Supplementary Material.

MD simulations were performed using the LAMMPS code [51] with a time step of 2 fs. Periodic boundary conditions (PBC) were applied for all directions to avoid surface effects, and the cell dimensions, angles, and individual atomic positions were allowed to fully relax during the simulations. The Nosé-Hoover thermostat and barostat [52, 53] were used to control the temperature and pressure, respectively. To avoid unstable atomic positions near grain boundaries and wire-matrix interfaces, each cell was relaxed at 0 K using the conjugate gradient method. Then, further relaxation was performed in an isobaric-isothermal (*NPT*) ensemble at zero pressure and the respective temperatures.

Phase transformation temperatures of the generated composite cells were investigated in an *NPT* ensemble at zero pressure. The cells were cooled from 400 K to 10 K and heated back to 400 K using cooling and heating rates of ±5 K/ps. Stress-controlled uniaxial cyclic tensile simulations along the axis of the wire were also performed in an *NPT* ensemble from zero to the selected maximum stress, and vice versa, at a stress rate of 6.25 MPa/ps. During the tensile simulations, the pressure perpendicular to the loading axis was controlled to be zero and the cell angles lying on that plane were allowed to change to reduce mechanical constraints caused by the PBCs. Young's moduli were determined by a linear fit of the loading curves from 0 to 300 MPa. The hysteresis of the stress-strain curves was calculated by integrating the area between the loading and unloading curves. To examine the effects of the size and loading rate on the overall results, independent simulations using cells with larger dimensions (maintaining the volume fraction of the wire) and with different loading rates were performed, and the results are provided in the Supplementary Material.

The polyhedral template matching method [54] implemented in OVITO [55] was used to visualize the occurrence of the martensitic phase transformation in the SMA matrix. It was confirmed [31, 33] that this method well distinguishes the bcc-like ordered B2 (austenite) and the monoclinic B19' (martensite) structures. Based on this method, the blue, red and gray atoms depicted in the figures of the atomic configurations represent austenite, martensite, and undetermined structures (such as grain boundaries, wire-matrix interface, amorphous region, and thermal noise). The austenite-martensite (*A-M*) phase boundary area calculation procedure carried out using OVITO is provided in the Supplementary Material. The dislocation extraction algorithm (DXA) [56] in OVITO was used to visualize and acquire detailed information on the evolution of the dislocations. The level of local plastic deformation was analyzed by visualizing the von-Mises local shear invariant [57] of each atom using OVITO.

## 3. Results and discussion

Prior to the cyclic tensile simulations, temperature-controlled MD runs were performed without any external stress. This process enables the detection of the phase transformation temperatures of the composite material that can be used for the selection of temperature ranges for further mechanical loading investigations. The martensite start ($M_s$) and the austenite finish ($A_f$) temperature of the composite material can be measured via identification of the sudden increase in the atomic volume due to the phase transformation from austenite (B2) to martensite (B19'), and vice-versa [45, 58]. The $M_s$ and $A_f$ temperatures were determined to be 190 and 260 K, respectively (cf. Supplementary Fig. S9a).

### *3.1. The origin of the peculiar mechanical behavior of nanowire+SMA composites*

We first examine the mechanical response of the composite and the evolution of its microstructure under cyclic tensile loading for one temperature, maximum stress, and dislocation density (other conditions are discussed in the following subsections). Specifically, four cycles at 300 K (40 K above $A_f$) up to a stress of 2450 MPa were performed. Figures 3a-b show the dislocation loops implanted into the nanowire before mechanical loading. During the initial relaxation at 300 K and zero-pressure for 16 ps, the vacancy loops relaxed into dislocation lines with a density of 0.04 nm$^{-2}$.

In the first cycle, a pseudoelastic behavior with a maximum deformation strain of about 8% is observed (black curve in Fig. 3c), albeit with an unusually large irrecoverable strain as compared to nanocrystalline SMAs without a nanowire [29]. As the cycling progresses, the loading curves appear to maintain linearity for over 2% strain in contrast to the first loading curve that exhibits linearity for less than 0.5% strain. Furthermore, the irrecoverable strains substantially decrease after the first cycle to values less than 0.6% despite a maximum deformation strain of over 4% during each cycle. For these reasons, we can conclude that the stress-strain responses from the second cycle onward can be classified as quasi-linear elastic with an accompanying reduction in hysteresis (from 58.6 to 10.0 MPa) and Young's modulus (from 68.9 to 49.7 GPa), similar to the experimental results with slightly larger (9.5%) pre-strains [15, 23]. In Fig. 3c, the change of the slopes in the middle of the loading curves indicated with the orange arrows occurs due to the completed martensitic transformation and the subsequent elastic straining of the fully transformed martensite phase in the NiTi SMA region. Apart from the slim hysteresis that accompanies quasi-linear elasticity as elucidated in the Introduction, the reduction of hysteresis with the number of loading cycles is a commonly observed phenomenon induced by pre-training (cyclic deformation) that is utilized before the final application to stabilize the hysteretic response of SMAs [59, 60].

To explain the origin of the peculiar dependence of the stress-strain curves, we have analyzed the fraction of retained martensite formed in each cycle. Figure 3d shows snapshots of the atomic configurations at zero and maximum stress. At the maximum stress of all four cycles, the SMA matrix is almost fully transformed to martensite. Upon unloading, 32.9% of the martensite is retained after the first cycle which progressively increases to 45.1% after the fourth cycle. As was demonstrated in a previous work [61], the reduction of Young's modulus can be attributed to the occurrence of retained martensite and the resulting availability of austenite-martensite (*A-M*) phase boundaries in the unloaded state.

The amount of retained martensite formed in the SMA matrix after a single tensile cycle is rather high considering the test temperature (40 K above $A_f$). Consequently, the material sample experiences a significant permanent change of its length as confirmed by the irrecoverable strain of 3.1%. This behavior can be traced back to slip occurring in the nanowire. The atomic local shear strain in Fig. 3e clearly reveals the occurrence of slip within the nanowire due to the movement of dislocations. We can conclude that the plastic deformation exhibited by the nanowire hinders the reverse transformation of the SMA matrix from martensite to austenite and thus assists in the stabilization of retained martensite, as was suggested by previous studies [15, 62]. The retained martensite is also related to the peculiar quasi-elastic behavior of the composite but in a rather subtle way as discussed next.

The microstructural origin of quasi-linear elasticity can be explicitly observed by inspection of the microstructural evolution during the present MD simulations (Fig. 4). In the top row (Figs. 4a-c), the austenite (blue) and martensite (red) regions at zero-stress (before the third cycle) are visualized from different perspectives. Below in Figs. 4d-l, the local lattice orientation (LLO) analysis [54] has been applied to provide additional, very instructive information on the microstructure development; specifically, for the states at zero-stress (d-f), at the maximum stress after the third cycle (g-i), and again at zero-stress after removal of the load (j-l). The different colors represent differently oriented grains/variants, and we have added white arrows to highlight the minor martensite variants which are of great relevance. It can be seen that these variants disappear on application of the load and reappear when the load is removed. These results comply with the explanation based on the Landau free energy landscapes [27, 28] mentioned in the Introduction. That means, quasi-linear elasticity and slim hysteresis for the pre-strained nanowire + SMA composite occur due to a defect-regulated nucleation and growth of martensite domains, which render the sharp, first-order martensitic transformation into a continuous process [15, 23].

We have performed similar simulations but with different loading rates and for larger cells to verify that the findings are independent of time and length scale constraints. The corresponding results (Supplementary Figs. S11-15) show the same changes in the stress-strain curves and properties.

### 3.2. Influence of dislocation density on the mechanical response

To determine the impact of the level of plastic deformation and with this to mimic varying defect conditions induced by the experimental synthesis, further simulations with different dislocation densities have been performed. About 0.02 and 0.09 nm$^{-2}$ of dislocations were introduced in the nanowire region before the mechanical loading (Figs. 5a-d; previous subsection: 0.04 nm$^{-2}$). Other conditions (i.e., cell dimensions, temperature, loading rate, number of cycles, and maximum stress) were kept the same as in Sec. 3.1.

Figures 5e-f show the resulting stress-strain curves and we can observe drastically different responses depending on the dislocation density. Likewise, significant differences in the slip behavior are apparent, Fig. 5g vs. Fig. 5h. The stress-strain response of the composite with the lower dislocation density (Fig. 5e) exhibits pseudoelastic behavior in all cycles, although Young's modulus reduces quite significantly from the first to the second cycle. This characteristic can be attributed to a too small fraction of retained martensite in the unloaded state and the resultant small area of *A-M* phase boundaries as shown in Fig. 5i. As explained above, the amount of retained martensite in the SMA region is enhanced by the plastic deformation of the nanowire region. However, not much slip for the lower dislocation density configuration could take place because the dislocation loops were initially located near the wire-matrix interface and could easily annihilate. Hence, hardly any retained martensite develops as shown in Fig. 5i. Similar results are obtained for a pristine nanowire NiTi composite without dislocations (Supplementary Fig. S16).

For the composite with the higher dislocation density, an evidently slim hysteresis and reduced Young's modulus (from 67.4 GPa to 41.0 GPa) can be observed after the first cycle, but Young's modulus significantly increases with further cycles (from 41.0 GPa to 56.4 GPa). This characteristic can be attributed to a very large martensite fraction in the unloaded state which increases further with increasing number of cycles as shown in Fig. 5j. The availability of many active slip systems obviously enhances the amount of retained martensite formed in the SMA region. However, if the retained martensite occupies a too large fraction of the SMA region, the area of *A-M* phase boundaries reduces and quasi-linear elasticity fades away. In a situation where the martensite occupies the majority of the SMA region, the deformation of the composite is in a regime of linear elasticity without any martensitic transformation. The stress-strain curve of the fourth cycle (green curve in Fig. 5f) indeed shows an almost usual linear elastic behavior.

The results presented so far firmly establish the relation between retained martensite and the mechanical response of the Nb nanowire + NiTi SMA composite. Plastic deformation of the nanowire plays a critical role in the evolution of retained martensite and is thereby responsible for the unusual mechanical properties of such composite materials. Especially, a moderate level of pre-strain is sufficient to trigger these unusual properties (experimentally observed for a pre-strain of 9 − 10% [15, 23]) due to the presence of nanowires, which in an experimental setting will likely contain enough dislocations to induce plastic deformation. As a counter-example, a nanowire-free NiTi SMA requires higher levels of pre-strain (experimental pre-strains greater than 27% [23, 27]) for a manifestation of quasi-linear elasticity and reduction in Young's modulus since the nanowire which could assist the formation of retained martensite is absent.

### *3.3. Influence of pre-straining on the mechanical response*

Additional simulations have been carried out to quantify the effect of varying degrees of pre-strain, controlled by different maximum stresses and numbers of cycles. The dislocation density was set to 0.04 nm$^{-2}$ and the temperature to 300 K, and the maximum stress varied from 1450 to 2450 MPa in intervals of 250 MPa. Previous experimental results [23] revealed that pseudoelasticity and a large hysteresis are maintained for low pre-straining (4%) whereas quasi-linear elasticity and a significant reduction in hysteresis occur for large pre-straining (9.5%). Our simulations exhibit similar strains from 5.6 to 7.9%, produced in the first cycle as seen in Figs. 6a-e.

Only the stress-strain curves for the higher maximum stresses, 1950, 2200 and 2450 MPa (Figs. 6c-e), show quasi-linear elasticity in the succeeding cycles, whereas for the lower maximum stresses (1450 and 1700 MPa) mostly a pseudo-elastic behavior is observed. This can be attributed to an insufficient amount of retained martensite occurring for the lower maximum stresses as shown in Fig. 6f. Nonetheless, a decrease in Young's modulus and hysteresis can be observed for every stress condition (Figs. 6g-h). These results reveal that the stress-strain response sensitively depends on the amount of strain and retained martensite produced during the cyclic loading.

To examine how the number of cycles affects the outcome for the lower stress conditions, we have increased the number of cycles up to 9 for the stress condition of 1450 MPa. The progressive development of the stress-strain curves and atomic configurations is shown in Figs. 7a-b. The $8^{th}$ and $9^{th}$ cycles markedly show quasi-linear elasticity and a noticeably small hysteresis and Young's modulus with a significant retained martensite fraction of over 20%. The total applied strain is 6.2% after the $7^{th}$ cycle (last cycle before the appearance of quasi-linear elasticity) which is among the lower levels of accumulated strains. This suggests that even a low pre-strain cycling can induce changes in the properties and hints us at the irrecoverable strain as a possible descriptor of the relevant physics. Irrecoverable strain is a straightforwardly measurable experimental quantity during a cyclic tensile test and its amount is known to be closely related to the retained martensite formed in SMAs [63]. We will elaborate on and quantify this relation in Sec. 3.5.

### 3.4. Influence of temperature on the evolution of microstructure and mechanical response

In conventional SMAs, the phase transformation temperatures (i.e., the $M_s$ and $A_f$ temperatures) are tunable factors by selecting appropriate compositions. Considering this possibility, additional simulations with varying temperatures have been performed to understand the influence of temperature on the transformation behavior and mechanical properties.

The corresponding MD runs were performed with a maximum stress of 2450 MPa for four cycles at temperatures of 240, 260, 280, 300, 320 and 340 K. After the initial relaxation at each designated temperature, the dislocation densities in the nanowire region were at a similar level (0.04, 0.03, 0.04, 0.04, 0.04, and 0.03 nm$^{-2}$ for the respective temperatures). However, a clear difference was observed in the stress-strain response in different temperature regions as shown in Figs. 8a-f. At relatively high temperatures (320 and 340 K), pseudoelasticity is mostly observed. At temperatures below and equal to $A_f$ (260 K), but higher than $M_s$ (190 K), i.e., 240 and 260 K, a nearly linear elastic behavior appears after the second cycle due to the deformation of the SMA region with an almost fully martensitic matrix. At moderate temperatures, i.e., 280 and 300 K, quasi-linear elasticity appears after the first cycle due to the previously discussed deformation mechanism including martensitic nanodomains.

### 3.5. Correlation of cumulative irrecoverable strain and Young's modulus

To verify our idea that irrecoverable strain provides a useful descriptor, we have carefully analyzed all of our MD data in search for the relevant correlations. Figure 9 compiles the most important results for the various stress and temperature conditions. The cumulative irrecoverable strain is the common *x*-axis and, on the y-axis, the relevant quantities resulting from the cyclic tensile simulations are captured.

Examination of the phase fractions (Fig. 9a) reveals a steady increase of retained martensite beyond 1% cumulative irrecoverable strain. The gradually increasing retained martensite fraction promotes a *continuous* martensitic transformation and consequently induces quasi-linear elasticity. This statement is well supported by the observed decrease of the hysteresis (Fig. 9b) along with the increase of the cumulative irrecoverable strain. Although we cannot quantitatively relate certain hysteresis values uniquely to quasi-linear elasticity because of the dependence on the details of the specific stress-strain curve, qualitatively we can state that larger hysteresis values are indicative of pseudoelasticity while smaller values of quasi-linear elasticity.

Young's modulus (Fig. 9c) exhibits a negative slope below a cumulative irrecoverable strain of 2.5% (indicated by the black dashed vertical line) and a positive slope beyond this strain. A similar phenomenon was reported [23] for highly pre-strained $Ni_{50.8}Ti_{49.2}$. The decrease of Young's modulus was attributed to a continued increase of *A-M* phase boundaries [61], and the increase of Young's modulus was attributed to the formation of amorphous regions [64]. The *A-M* phase boundary area in our simulations indeed exhibits a steep increase with increasing cumulative irrecoverable strain (Fig. 9d) for the strains where Young's modulus shows a decrease. When the *A-M* phase boundary area is about to reach its maximum (indicated by the yellow shaded ellipse), the minimum in Young's modulus is observed. The difference between the actual maximum of the fitted Gaussian in Fig. 9d and the minimum of Young's modulus in Fig. 9c is likely due to inaccuracies in the identification of surfaces and/or interfaces (caused by roughness, over/under-projection, etc.). Nevertheless, this correlation between Young's modulus and the *A-M* phase boundary area mediated by the cumulative irrecoverable strain is in good agreement with the previous results reported by Zhang *et al.* [61].

For the strains where the *A-M* phase boundary area is at its maximum and then decreases again, Young's modulus shows an increasing trend. In this region the softening effect of the *A-M* phase boundary area is saturated and the continuing increase of the phase fraction of the stiffer martensite (Fig. 9a) drives the increase of Young's modulus. In fact, the increase is so strong that Young's moduli beyond a cumulative irrecoverable strain of 4.5% are observed to be greater than their initial values. This can be attributed to the dominance of martensite which then occupies 50% of the matrix.

Our analysis thus demonstrates the suitability of the cumulative irrecoverable strain as a descriptor to correlate and monitor properties, especially Young's modulus. In general, the values of Young's modulus and phase fraction could vary depending on grain size, volume fraction of the nanowire, composition of the matrix, etc. However, we expect that the cumulative irrecoverable strain can still be used to monitor the appropriate level of pre-straining by fine tuning the details for each material.

## 4. Conclusion

We have provided an atomic-scale understanding of the unusual mechanical properties of Nb nanowire + NiTi SMA composites via MD simulations based on a newly developed interatomic potential. Through the simulations, the origin of quasi-linear elasticity, slim hysteresis, and reduction in Young's modulus occurring for pre-strained composites has been revealed by analyzing and visualizing the details of the atomic-scale processes. Our work has further revealed that these mechanical properties can be tuned by employing cyclic pre-straining in appropriate temperature ranges. The results can be summarized as follows:

1) The developed interatomic potential for the Nb-Ni-Ti ternary system faithfully reproduces not only various fundamental physical properties but also the experimentally observed unusual mechanical response and features of a pre-strained Nb nanowire embedded in a NiTi SMA matrix.

2) It is the retained martensite and nanodomains of different variants developed through pre-straining, which change the sharp, first-order martensitic transformation into a continuous transformation process and which produce *A-M* phase boundaries that are responsible for quasi-linear elasticity, slim hysteresis, and varying values of Young's modulus.

3) The occurrence of plastic deformation of Nb nanowires within the SMA matrix during pre-straining promotes the formation of nanodomains of martensite, thus inducing quasi-linear elasticity and a change in Young's modulus earlier than observed for conventional SMAs without nanowires.

4) To control and obtain a desired Young's modulus for a composite, cyclic pre-straining can be employed at appropriate temperatures, while simultaneously monitoring the Young's modulus with the help of the cumulative irrecoverable strain that is closely associated with the microstructure evolution and straightforwardly attainable from stress-strain curves.

**Data Availability**

The data that support the finding of this study is available from the corresponding author (email: wonsko@ulsan.ac.kr) upon reasonable request.

**Acknowledgements**

We thank Alexander Stukowski for providing us with his tool for dislocation loop creation. This project has received funding from the European Research Council (ERC) under the European Union's Horizon 2020 research and innovation programme (grant agreement No 865855). This research was also supported by the National Research Foundation of Korea (NRF) funded by Ministry of Science and ICT (Grant No. NRF-2019M3D1A1079214, NRF-2019M3E6A1103984, and NRF-2019R1F1A1040393). B.G. also acknowledges the support by the Stuttgart Center for Simulation Science (SimTech). W.-S.K. also acknowledges the support by the National Supercomputing Center with supercomputing resources (Grant No. KSC-2020-CRE-0332).

**Appendix A: Optimization of the Ni-Ti-Nb 2NN MEAM potential**

To describe a pure element using the 2NN MEAM formalism, 14 independent parameters are necessary, namely, the cohesive energy ($E_c$), the equilibrium nearest-neighbor distance ($r_e$), the bulk modulus of the reference structure ($B$), the embedding function parameter ($A$), the decay lengths ($\beta_0, \beta_1, ... \beta_3$), the weighting factors ($t_1$, $t_2$, and $t_3$), the many-body screening parameters ($C_{\min}$ and $C_{\max}$), and one adjustable parameter ($d$). Table 1 lists the parameters used for pure Ni, Ti, and Nb. The unary parameters of pure Ni and Ti were adopted from our previous work [45]. The optimization of the parameters for pure Nb was carried out in the present work by minimizing errors between the forces between atoms and structural energies obtained by the DFT calculations and temporary potential parameters considering atomic configurations under various situations. For example, configurations in perfect and imperfect (in presence of vacancies and thermal vibrations) states of bcc, fcc, and hcp structures and configurations of liquid structures were included in the target database.

Apart from the 14 unary parameters for each pure element, 13 independent parameters are required to describe a binary system. In addition to $r_e$, $B$, and $d$ which were used in the unary case, $\Delta E_f$ (enthalpy of formation of the reference structure), $\rho_0$ (electron density scaling factor), and four $C_{\min}$ and four $C_{\max}$ are necessary. To describe a ternary system, on the other hand, only 6 additional parameters (three $C_{\min}$ and three $C_{\max}$) are needed. Table 2 and 3 list the binary and ternary parameters used in this work, respectively. Again, the optimized binary NiTi parameters were used from our previous work [45]. For optimizing the binary Nb-Ni, Nb-Ti, and ternary Ni-Ti-Nb parameters, the procedure from our previous work [45] was employed, i.e., optimization focusing only on the physical properties of intermetallic compounds. The properties of the solid solutions such as the solute-solute binding energy, solute migration energy, vacancy-solute binding energy, and dilute heat of the solution were used to examine the transferability of the developed alloy potentials. The performance of the developed potentials was evaluated through comparison of various physical properties obtained from previous experiments and the present DFT calculations. For accuracy tests and transferability of the developed potential, we refer to the Supplementary Material.

A radial cutoff distance of 5 Å was used for all simulations throughout this work as this choice well reproduced physical properties and the martensitic transformation of NiTi SMA at good computational efficiency.

**Appendix B: Computational details of the first-principles DFT calculations**

The present DFT calculations were performed using the VASP code [65, 66] with the implementation of the projector-augmented wave method [67, 68] within the Perdew-Burke-Ernzerhof generalized-gradient approximation [69] for the exchange-correlation functional. A cutoff energy of 400 eV for the plane-wave basis set was used. A $\Gamma$-centered $k$-point mesh of 21×21×21 was selected for the bcc unit cell with a single atom and the similar $k$-point density was implemented for other structures. The Methfessel-Paxton smearing method [70] with a width of 0.1 eV was applied. For Ni-rich structures and intermetallic compounds, magnetism was considered via spin-polarized calculations. The Birch-Murnaghan equation of state [71, 72] was used to calculate the lattice constants and bulk moduli.

Defect properties were calculated by relaxing atomic positions at fixed volume and cell shape with the convergence criteria of forces and energies set as $10^{-2}$ eV/Å and $10^{-6}$ eV, respectively. The nudged elastic band method [73, 74] was utilized to calculate the vacancy migration energy. Surface energies were calculated using slabs having thickness of $20-25$ Å and a vacuum region of ~10 Å. For phonon calculations, the Phonopy code [75] was used based on the direct force constant approach with a bcc supercell of 250 atoms and convergence criteria of forces and energies set as $10^{-4}$ eV/Å and $10^{-8}$ eV, respectively.

For atomic configurations at finite temperatures, the concept of the UP-TILD method [76] was followed, performing two-step DFT calculations. Snapshots of *ab initio* MD simulations for 2000 steps with 2 fs time step and lenient convergence criteria were taken in the first step. Subsequently, DFT calculations using higher cutoff energies and denser $k$-point mesh of the previously taken snapshots were carried out to obtain accurate forces and energies.

TABLE 1. Optimized 2NN MEAM potential parameter sets for pure Nb, Ni, and Ti. The following properties are dimensionful: the cohesive energy $E_c$ (eV/atom), the equilibrium nearest-neighbor distance $r_e$ (Å), and the bulk modulus $B$ ($10^{12}$ dyne/cm²). The reference structures are bcc Nb, fcc Ni, and bcc Ti.

|    | $E_c$ | $r_e$ | $B$ | $A$ | $\beta^{(0)}$ | $\beta^{(1)}$ | $\beta^{(2)}$ | $\beta^{(3)}$ | $t^{(1)}$ | $t^{(2)}$ | $t^{(3)}$ | $C_{min}$ | $C_{max}$ | $d$ |
|---|---|---|---|---|---|---|---|---|---|---|---|---|---|---|
| Nb | 7.47 | 2.860 | 1.7624 | 0.74 | 5.11 | 4.86 | 3.50 | 4.42 | 1.24 | 2.87 | 5.12 | 0.49 | 2.49 | 0.00 |
| Ni[a] | 4.45 | 2.490 | 1.8586 | 0.79 | 2.48 | 1.94 | 3.46 | 2.56 | 2.84 | -1.20 | 2.49 | 0.95 | 1.75 | 0.05 |
| Ti[a] | 4.75 | 2.850 | 1.0735 | 0.24 | 2.20 | 3.00 | 4.00 | 3.00 | -18.0 | -32.0 | -44.0 | 0.25 | 1.58 | 0.00 |

[a] Ref. [45].

TABLE 2. Optimized 2NN MEAM potential parameter set for the binary Nb-Ni, Nb-Ti, and Ni-Ti systems. The following quantities are dimensionful: the enthalpy of formation of the reference structure $\Delta E_f$ (eV/atom), the equilibrium nearest-neighbor distance $r_e$ (Å), and the bulk modulus $B$ ($10^{12}$ dyne/cm²).

| Parameter | Nb-Ni | Nb-Ti | Ni-Ti[a] |
|---|---|---|---|
| Reference structure | Cu₃Au type Nb₃Ni | CsCl type NbTi | CsCl type NiTi |
| $\Delta E_f$ | 0.00 | -0.03 | -0.36 |
| $r_e$ | 2.836 | 2.834 | 2.612 |
| $B$ | 1.9226 | 1.0462 | 1.2818 |
| $d$ | $0.75d^{Nb}+0.25d^{Ni}$ | $0.5d^{Nb}+0.5d^{Ti}$ | $0.5d^{Ni}+0.5d^{Ti}$ |
| $\rho_0^A : \rho_0^B$ | 1 : 1 | 1 : 1 | 1 : 1 |
| $C_{min}^{A-B-A}$ | 0.80 | 1.00 | 0.25 |
| $C_{min}^{B-A-B}$ | 0.09 | 0.09 | 0.09 |
| $C_{min}^{A-A-B}$ | 0.09 | 1.00 | 0.49 |
| $C_{min}^{B-B-A}$ | 0.49 | 0.90 | 1.60 |
| $C_{max}^{A-B-A}$ | 1.44 | 2.80 | 1.70 |
| $C_{max}^{B-A-B}$ | 1.44 | 2.80 | 1.70 |
| $C_{max}^{A-A-B}$ | 2.00 | 1.44 | 1.40 |
| $C_{max}^{B-B-A}$ | 1.44 | 1.44 | 1.70 |

[a] Ref. [45].

**TABLE 3**. Optimized 2NN MEAM potential parameter set ($C_{max}$ and $C_{min}$) for the ternary Nb-Ni-Ti system.

| Parameter | Nb-Ni-Ti |
|---|---|
| $C_{min}^{Nb-Ti-Ni}$ | 0.09 |
| $C_{min}^{Nb-Ni-Ti}$ | 0.90 |
| $C_{min}^{Ni-Nb-Ti}$ | 0.90 |
| $C_{max}^{Nb-Ti-Ni}$ | 1.44 |
| $C_{max}^{Nb-Ni-Ti}$ | 1.44 |
| $C_{max}^{Ni-Nb-Ti}$ | 2.80 |

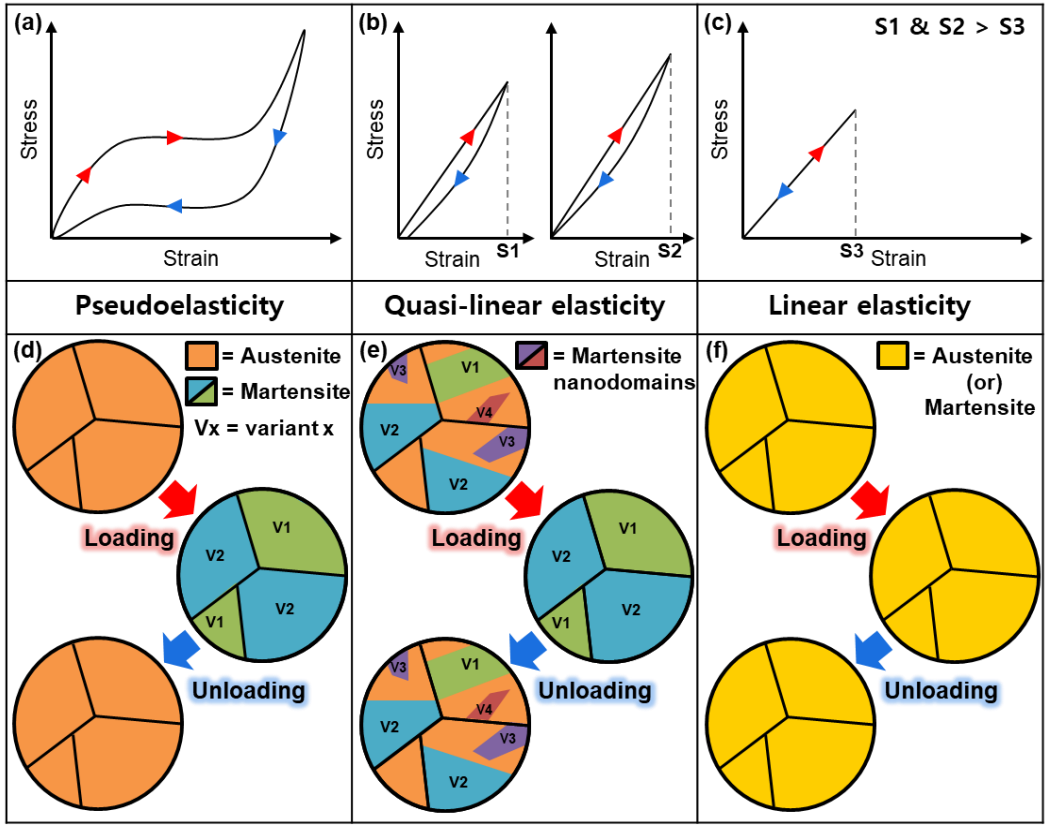

**Fig. 1.** (a-c) Schematic stress-strain curves representing pseudoelasticity, quasi-linear elasticity, and linear elasticity, respectively. S1 to S3 indicate the maximum applied strains in their respective curves. (d-f) Microstructure changes during the deformation cycle for pseudoelastic, quasi-linear elastic, and linear elastic cases, respectively. Legends for the microstructure are provided in the figure.

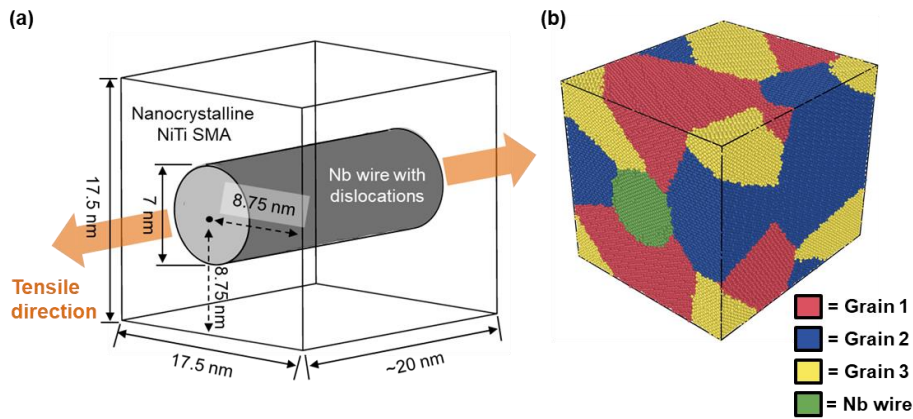

**Fig. 2.** (a) Schematic diagram of the Nb nanowire + NiTi SMA composite cell with dimensions of ~20×17.5×17.5 nm and wire diameter of 7 nm. The tensile direction is along the axis of the wire as indicated by the orange arrows. (b) Utilized nanocrystalline cell composed of 3 NiTi grains with the colors (red, blue, and yellow) indicating different grain orientations (note that periodic boundary conditions apply). The Nb wire is colored green. Grain boundaries are not shown.

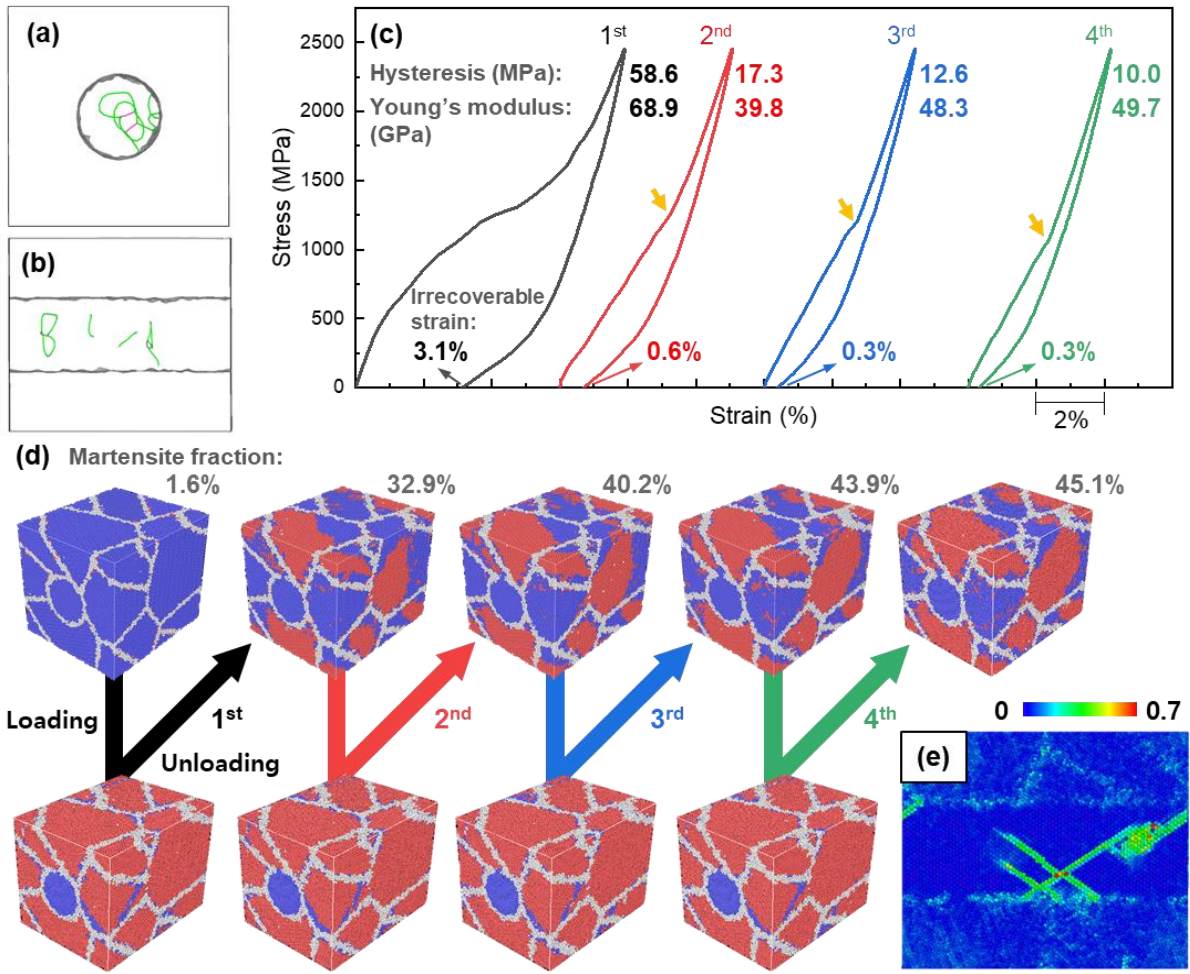

**Fig. 3.** (a) Front and (b) side views of the composite cell showing dislocations after the relaxation at 300 K (dislocation density of 0.04 nm$^{-2}$). Green lines correspond to $\frac{1}{2}$<111> and magenta lines to <100> dislocations. (c) Cyclic stress-strain curves of the dislocation containing Nb nanowire + NiTi SMA composite simulated at 300 K. Corresponding hysteresis (in MPa), Young's modulus (in GPa), and irrecoverable strain values are indicated beside each curve. Orange arrows indicate the changes in the slopes due to the completion of the martensite transformation. (d) Evolution of the microstructure during each cyclic loading. The blue atoms correspond to the austenite (B2) structure, red atoms to the martensite (B19') structure, and gray atoms represent undetermined structures (grain boundary, wire-matrix interface, amorphous region, and thermal noise). The fraction of martensite at each unloaded state is provided beside each cell. (e) Cross-sectional view of the composite cell visualizing the local shear strain [57].

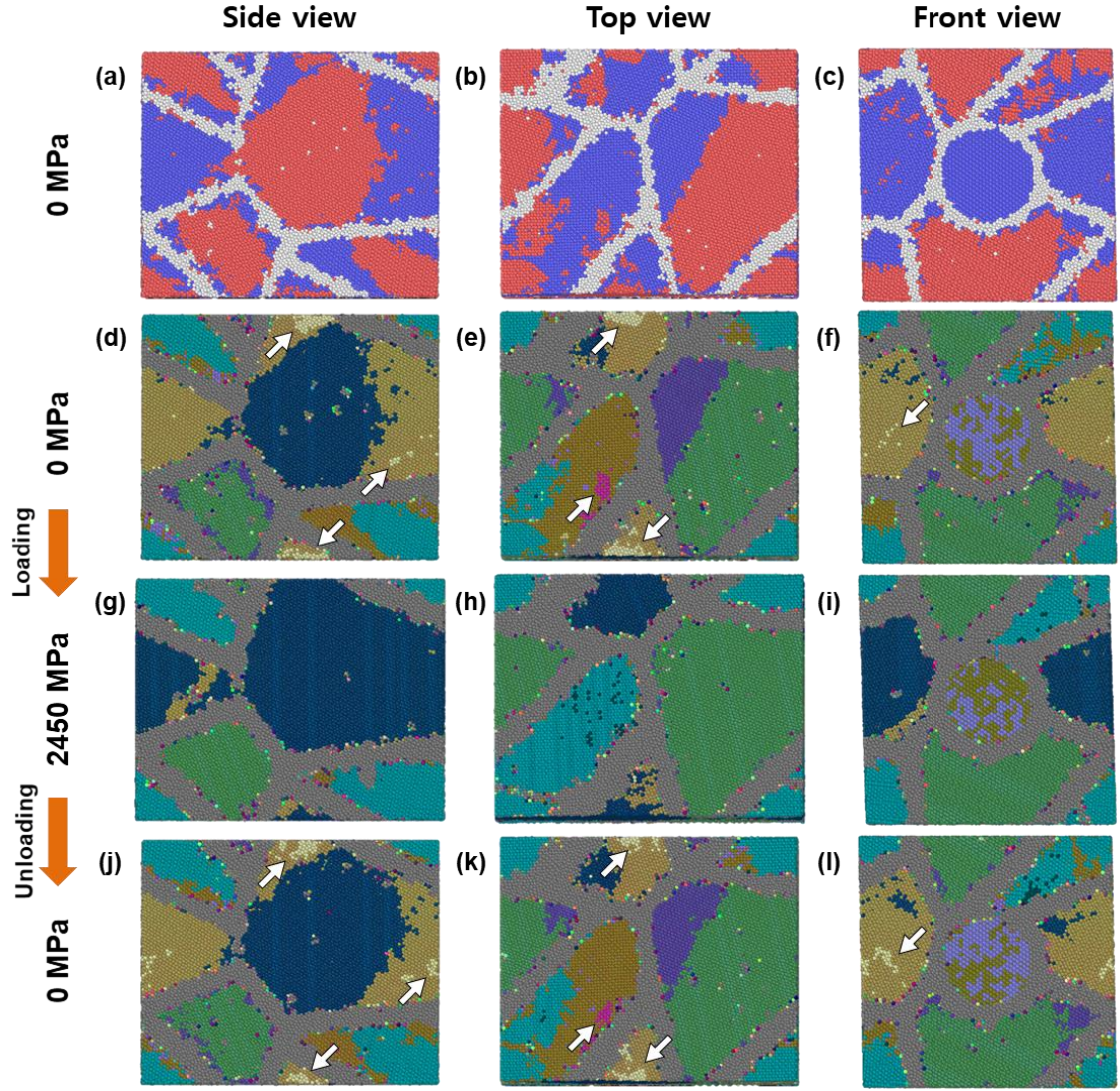

**Fig. 4.** (a-c) Side, top, and front view (in order) snapshots of atomic configurations of the Nb nanowire + NiTi SMA composite (dislocation density of 0.04 nm$^{-2}$) before the third cycle loading at 300 K. The blue atoms correspond to the austenite (B2) structure, red atoms to the martensite (B19') structure, and gray atoms represent undetermined structures (grain boundary, wire-matrix interface, amorphous region, and thermal noise). (d-l) Side, top, and front views of the same composite during the third cycle with local lattice orientation (LLO) [54] applied. (d-f) correspond to the zero-stress condition before the third cycle, (g-i) correspond to the maximum stress condition (2450 MPa) during the third cycle, and (j-l) correspond to the zero-stress condition after the third cycle. The white arrows indicate the minor variants of martensite.

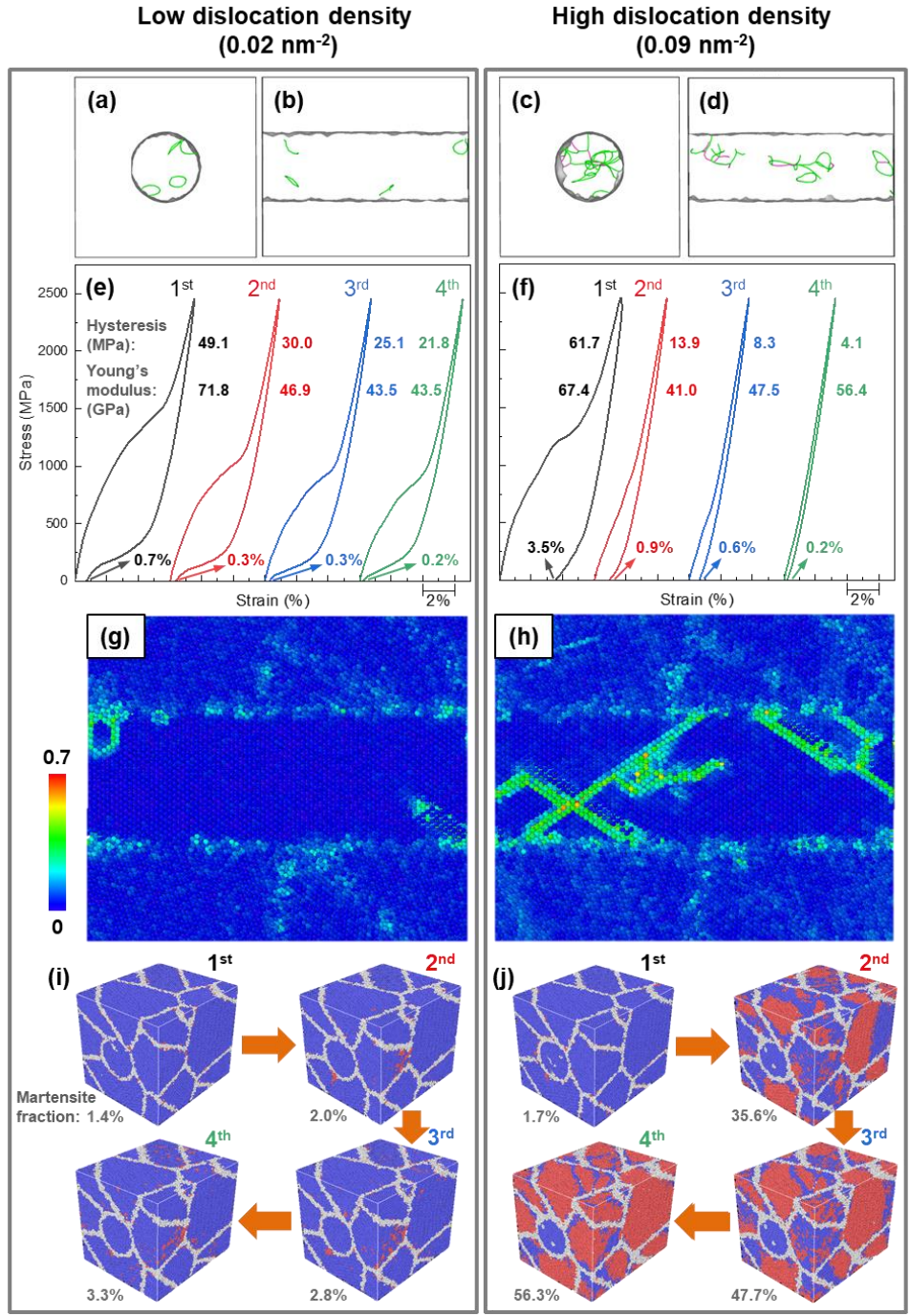

**Fig. 5.** (a-d) Front and side views of composite cells (20×17.5×17.5 nm) showing dislocations after the relaxation at 300 K. Green lines correspond to $\frac{1}{2}$<111> and magenta lines to <100> dislocations. (e-f) Cyclic stress-strain curves of the Nb nanowire + NiTi SMA composite with different dislocation densities at 300 K. Corresponding hysteresis (in MPa), Young's modulus (in GPa), and irrecoverable strain values are indicated beside each curve. (g-h) Cross-sectional view of the composite cells visualizing the local shear strain [57] after a single tensile cycle. (i-j) Atomic configurations before each loading. The blue atoms correspond to the austenite (B2) structure, red atoms to the martensite (B19') structure, and gray atoms represent undetermined structures (grain boundary, wire-matrix interface, amorphous region, and thermal noise). The fraction of martensite at each unloaded states is provided beside each cell. (a, b, e, g, and i) correspond to a cell with dislocation density of 0.02 nm$^{-2}$. (c, d, f, h, and j) correspond to a cell with dislocation density of 0.09 nm$^{-2}$.

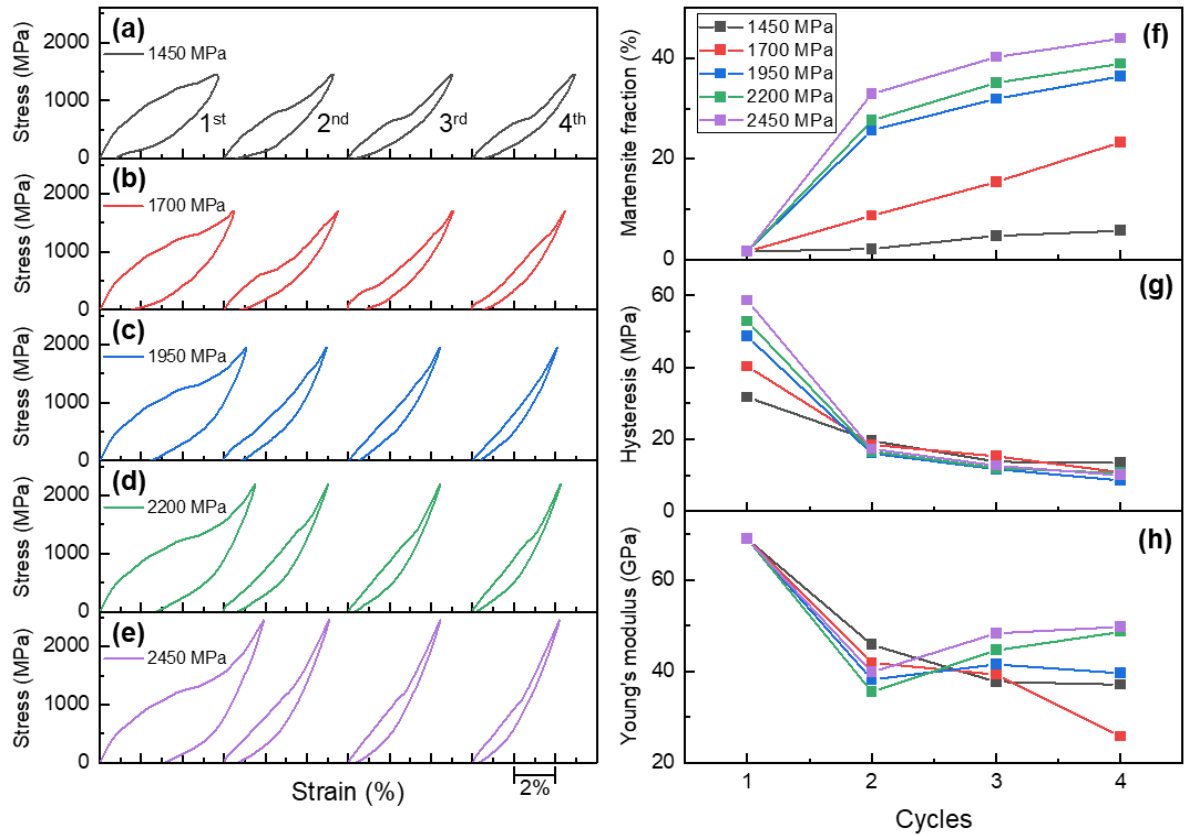

**Fig. 6.** (a-e) Cyclic stress-strain curves of the Nb nanowire + NiTi SMA composite (dislocation density of 0.04 nm$^{-2}$) simulated at 300 K with different maximum stresses. (f) Martensite fraction, (g) hysteresis, and (h) Young's modulus obtained during each cycle of the tensile simulations with different maximum stresses. Lines connecting the symbols in (f-h) are presented only as guide to the eyes.

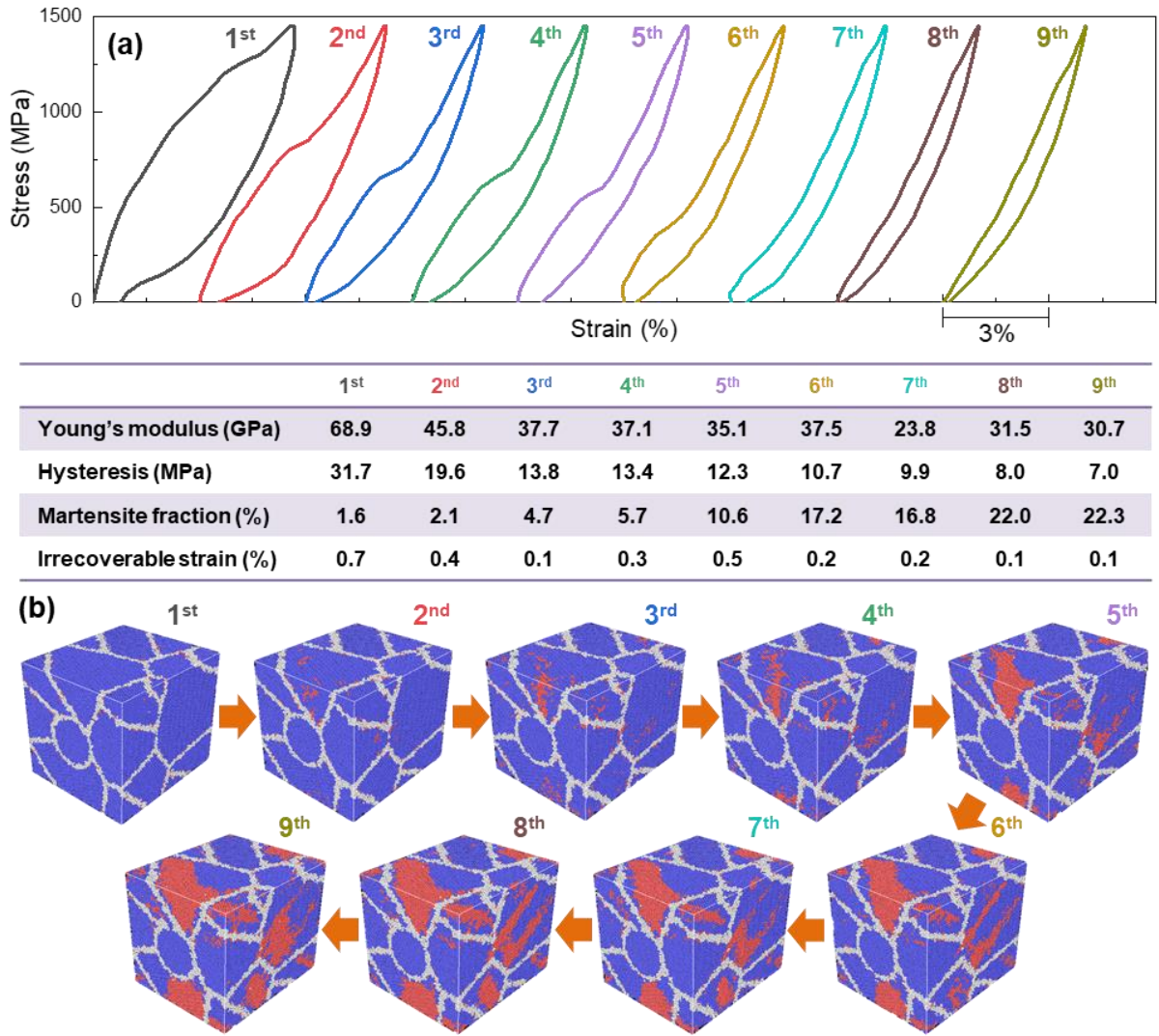

**Fig. 7.** (a) Cyclic stress-strain curves of the Nb nanowire + NiTi SMA composite (dislocation density of 0.04 nm$^{-2}$) simulated at 300 K with maximum stress of 1450 MPa for up to 9 cycles. Corresponding Young's modulus (in GPa), hysteresis (in MPa), martensite fraction before loading, and irrecoverable strain undergone during each cycle are provided in the table below. (b) Atomic configurations before each loading. The blue atoms correspond to the austenite (B2) structure, red atoms to the martensite (B19') structure, and gray atoms represent undetermined structures (grain boundary, wire-matrix interface, amorphous region, and thermal noise).

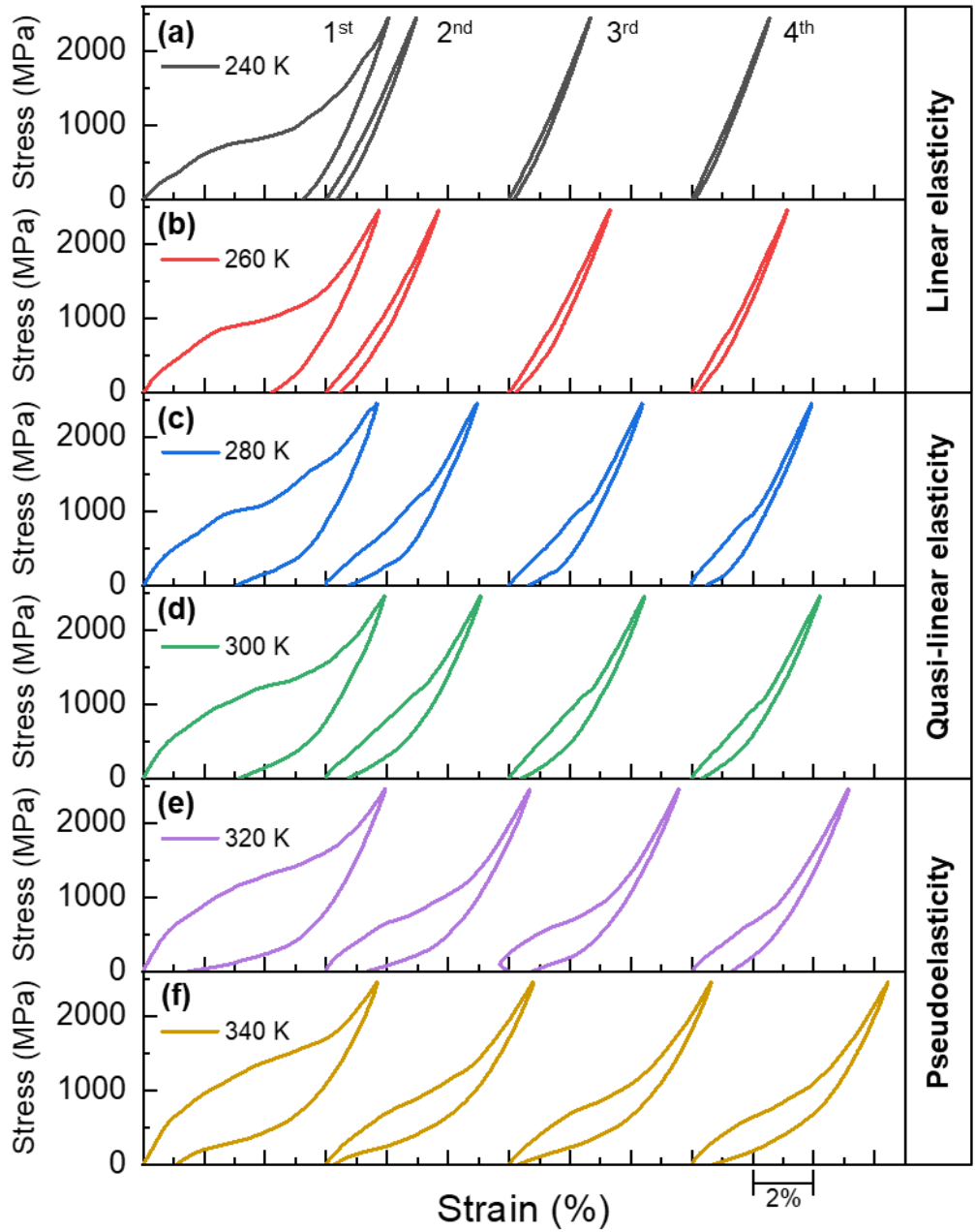

**Fig. 8.** (a-f) Cyclic stress-strain curves of the Nb nanowire + NiTi SMA composite simulated at different temperatures with maximum stress of 2450 MPa. The "elasticity labels" to the right indicate the dominant behavior of the stress-strain response at the respective temperature.

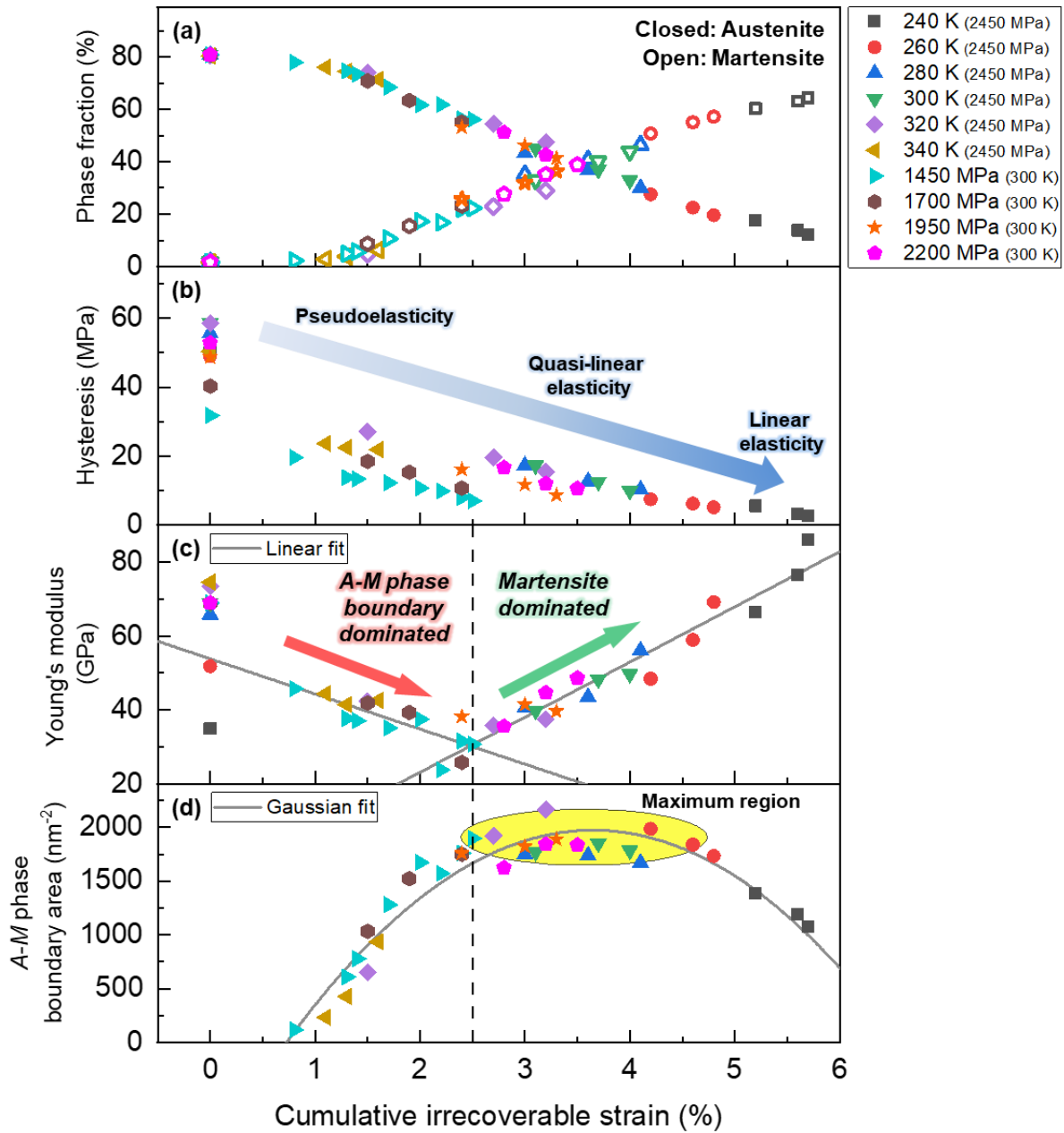

**Fig. 9.** (a) Phase fraction of austenite and martensite in the matrix, (b) hysteresis, (c) Young's modulus, and (c) *A-M* phase boundary area plotted against the corresponding cumulative irrecoverable strain obtained from cyclic tensile simulation at different temperatures and maximum stresses. The same color coding is used from (a) to (d). Closed and open symbols represent austenite and martensite fractions, respectively in (a). The black vertical dashed line at 2.5% cumulative irrecoverable strain in (c-d) indicates the change in the sign of the slope of Young's modulus. The maximum region of *A-M* phase boundary area in (d) is marked by a yellow ellipse.

# Supplementary Material to "Atomistic simulations of the deformation behavior of an Nb nanowire embedded in a NiTi shape memory alloy"

Jung Soo Lee [a], Won-Seok Ko [b,*], Blazej Grabowski [c]

[a] *Research Institute of Basic Sciences, University of Ulsan, Ulsan 44610, Republic of Korea*

[b] *School of Materials Science and Engineering, University of Ulsan, Ulsan 44610, Republic of Korea*

[c] *Institute for Materials Science, University of Stuttgart, Pfaffenwaldring 55, 70569 Stuttgart, Germany*

# S-1. Supplementary description

## S-1.1. Cell configurations for the investigation on the effect of dislocation density

Nanocrystalline B2 SMA ($Ni_{50}Ti_{50}$) cells with three grains were generated using the Voronoi construction method [1] at random positions with arbitrary crystallographic orientations for each grain. The atoms positioned within a cylinder with its axis lying at the center of the cell parallel to the direction with the longest edge were substituted with Nb atoms. The Nb nanowire was oriented in [110] direction along the axis of the wire. Cell dimensions and the diameter of the Nb wire were ~20×17.5×17.5 nm and 7 nm, respectively. The procedure proposed by Zepeda-Ruiz et al. [2] was adopted to generate dislocations within the Nb wire. Two dislocation loop conditions were imposed: (1) 12 vacancy loops having a diameter of 3.5 times the lattice parameter (0.33 nm), and (2) 44 vacancy loops having a diameter of 4 times the lattice parameter (0.33 nm) were randomly positioned within the single crystal bcc Nb cuboid with dimension ~20×7.5×7.3 nm. Subsequently, the conjugate gradient method was used to minimize the energy, relaxing the "vacancy platelets" to form dislocation loops and networks as shown in Fig. S1. Then, a cylinder with diameter of 7 nm was extracted and inserted into cylindrically hollow cuboidal nanocrystalline NiTi supercell with. After the preparation of the nanowire + SMA composite cell, relaxation was performed in an *NPT* ensemble at zero pressure and at 300 K. The number of atoms in the SMA and nanowire regions were around $3.9 \times 10^5$ and $4.2 \times 10^4$, respectively.

## S-1.2. Cell configurations for the investigation on the influence of the stress rate

The same procedure as mentioned in the previous section, S-1.1, was followed to generate the Nb nanowire + NiTi SMA composite cell. However, a different dislocation loop condition was used, producing 24 vacancy loops with a diameter of 4 times the lattice parameter (0.33 nm) in the Nb single crystal. After the preparation of the nanowire + SMA composite cell, relaxation was performed in an *NPT* ensemble at zero pressure and at 300 K. The number of atoms in the SMA and nanowire regions were as above, around $3.9 \times 10^5$ and $4.2 \times 10^4$, respectively.

## S-1.3. Cell configurations for the investigation on the influence of cell size

The same procedure as mentioned in section S-1.1 was followed to generate Nb nanowire + NiTi SMA composite cells of different sizes while maintaining the volume fraction of the wire (~12.5%). The number of grains, dimensions of the composite cell, and number of atoms are shown in table S1. The same dislocation loop condition as in section S1.2 was implemented. After the preparation of the nanowire + SMA composite cell, relaxations were performed in an *NPT* ensemble at zero pressure and at respective temperatures of austenite finish ($A_f$) +20 K.

## S-1.4. Austenite-martensite phase boundary area calculation method

The calculation of the austenite-martensite (*A-M*) phase boundary area was performed by taking the average of values obtained from two different methods. The average was taken as the final phase boundary area since the identification of grains/particles of austenite/martensite/undetermined structures slightly differs between the two methods when the sizes of grains/particles are small. The two methods are explained in the following with reference to the schematic in Fig. S2.

1) In the first method, the *A-M* phase boundary area is calculated by adding the austenite surface area ($A^s$) and martensite surface area ($M^s$), and subtracting the surface area of jointly identified austenite-martensite grains ($(A+M)^s$). It is mathematically expressed as follows:

$$A\text{-}M \text{ surface } = \frac{A^s + M^s - (A+M)^s}{2}.$$

The expression is divided by 2 since the phase boundary area is duplicated by considering $A^s$ and $M^s$.

2) The second method indirectly measures the *A-M* phase boundary area by considering the surface areas of undetermined structures. First, the martensite exclusive surface area ($M^{ex}$) is evaluated excluding the *A-M* phase boundary surface. This is obtained by adding the undetermined structure surface area ($U^s$) and martensite surface area ($M^s$), and subtracting the surface area of jointly identified undetermined structure-martensite grains ($(U+M)^s$) which is equivalent to the austenite surface area. The mathematical expression is:

$$M^{ex} = \frac{U^s + M^s - (U+M)^s}{2}.$$

Again, the expression is divided by 2 due to the duplication considering $U^s$ and $M^s$. Finally, the *A-M* phase boundary area is obtained by subtracting the martensite exclusive surface area ($M^{ex}$) from the martensite surface area ($M^s$):

$$A\text{-}M \text{ surface } = M^s - M^{ex}.$$

**S-1.5. Accuracy and transferability of the developed Ni-Ti-Nb potential**

Physical properties calculated using the developed interatomic potential for the Ni-Ti-Nb ternary system were compared to experimental and DFT results to evaluate the accuracy and transferability of the developed potential. All the MD simulations and molecular static simulations were performed with supercells with at least 4000 atoms using the LAMMPS code [3]. The Nosé-Hoover thermostat and barostat [4, 5] were used to control the temperature and pressure, respectively. A timestep of 2 fs was used and periodic boundary conditions in all three dimensions were applied except for the calculation of surface energies. For thermal properties, MD simulations were carried out in an isobaric-isothermal (*NPT*) ensemble at zero pressure. The melting temperature was calculated using the interface method. Subsequently, the enthalpy of melting and volume change upon melting were determined using the calculated melting temperature. For the molecular statics simulations at 0 K full relaxation of the cell dimensions and all atomic positions was performed.

### S-1.5.1. Physical properties of pure Nb

Figure S3 shows the scatter plots for energies and forces of pure Nb. Compared to the previous 2NN MEAM potential [6] (Lee *et al*. [6]), the developed potential exhibits a better correlation with the DFT values indicating its superior accuracy in the fitting. In table S2 the various bulk, elastic, defect, and surface properties of pure Nb at 0 K are shown and compared with reported experimental data, DFT calculations, and results based on the previous potential. The good reproducibility of these properties indicates the accuracy of fitting because they are closely related to atomic configurations used in the force matching process. Overall, the properties are well reproduced, showing results closer to those of experiments and DFT calculations than the previous potential by Lee *et al*. [6] for most values.

The transferability of the potential was examined through comparison of thermal properties such as the thermal expansion coefficient, specific heat, melting temperature, enthalpy of melting, and volume change upon melting. As shown in table S3, better agreement with experimental data is exhibited by the developed potential than by the previous potential [6]. Phonon spectra of bcc Nb are compared in Fig. S4. The overall dependence of the phonon branches of the developed potential is in good agreement with both the experimental and DFT data while those of the previous potential exhibit a significant underestimation of frequencies around the point *H*. A further transferability check was done by comparison of the generalized stacking fault energy (GSFE) curves of bcc Nb for <111> dislocations. As can be seen from Fig. S5, the GSFE curves of <111> dislocations on {110} and {112} slip planes obtained using the developed potential are closer to the DFT data than for the previous potential [6].

### S-1.5.2. Physical properties of binary Nb-Ni and Nb-Ti alloys

For the binary Nb-Ni and Nb-Ti subsystems, the physical properties of the solid solutions were analyzed to evaluate the performance and accuracy of the potential. Table S4 lists the properties of the solid solutions calculated at 0 K using the present potential along with DFT data. As can be observed from table S4, the solid solution related energies are satisfactorily reproduced in general. For some of the smaller values, i.e., weak interactions, the sign disagrees with the DFT prediction (positive sign = attractive interaction, negative sign = repulsive interaction). However, for the strong interactions, the same signs and similar values as in DFT are predicted by the potential.

Figure S6 shows the lattice constants of Nb-Ni (Ni-rich) fcc and Nb-Ti (Nb-rich) bcc solid solutions determined using the developed potential in comparison to experimental data. The calculated lattice constants in both cases follow well the increasing/decreasing trend of the experimentally measured data. To check the transferability to the liquid phase at high temperature, the enthalpy of mixing of the Nb-Ni and Nb-Ti liquid phases was calculated at 2800 K using the current potential and the CALPHAD method. As shown in Fig. S7, apart from small deviations from the CALPHAD based data, the overall trend throughout the compositions for both binary systems is satisfactorily described, indicating a good transferability of the developed binary potentials over a wide range of compositions.

Additionally, the properties (lattice constants, bulk moduli, and enthalpy of formation) of stable as well as hypothetical intermetallic compounds in the binary Nb-Ni and Nb-Ti systems were calculated and compared for further transferability evaluation. In table S5 the calculated properties along with experimental and DFT data are shown. For the Nb-Ni system, the properties of both the stable ($D0_a$-NbNi$_3$ and $D8_5$-Nb$_7$Ni$_6$) and hypothetical compounds obtained using the developed potential are well reproduced. Despite the lack of stable phases for the Nb-Ti system, the properties are reasonably reproduced. In Fig. S8 atomic snapshots obtained during the thermal loading simulation process for the stable compounds $D8_5$-Nb$_7$Ni$_6$ and $D0_a$-NbNi$_3$ are illustrated. The structures were well maintained for a wide range of temperatures, indicating the good quality of the binary Nb-Ni/Ti interactions in the developed potential.

### S-1.5.3. Physical properties of ternary Nb-Ni-Ti alloys

The evaluation for the ternary Nb-Ni-Ti system was based on the comparison with the present DFT calculations due to the lack of experimental information for this ternary system. The accuracy evaluation was carried out by comparing the solute-solute binding energy of two different types of solutes in the solid solutions of Nb-rich bcc, Ni-rich fcc, and Ti-rich hcp to the DFT results since these properties were used for the parameter optimization. The calculated energies are shown in Table S6. The current potential well reproduces the attractive or repulsive interactions of the solute atoms (indicated by positive or negative signs) as predicted by the DFT calculations.

The transferability evaluation was done by comparing the properties (lattice constants, bulk moduli, and enthalpy of formation) of hypothetical Cu$_2$MnAl-type ternary compounds. Table S7 lists the calculated properties of the compounds. In general, the calculated properties are in agreement with the DFT data. The calculated lattice constants are close to the DFT data while some deviation exists for the bulk moduli and enthalpy of formation.

In conclusion, the solid solutions as well as intermetallic compounds can be described reasonably well using the developed potential and we believe that it can successfully be applied for other atomistic simulations of Ni-Ti-Nb systems with acceptable consistency.

## S-2. Supplementary tables

**TABLE S1.** Dimensions (nm) of composite cells, number of grains in the NiTi SMA matrix, diameter of the Nb nanowire (nm), and number of Ni, Ti and Nb atoms in the composite cells.

|  | Cell dimensions (nm) | Number of grains | Wire diameter (nm) | Number of atoms Ni + Ti | Nb |
|---|---|---|---|---|---|
| Composite cell 1 | ~30×26×26 | 5 | 10.4 | ~1.3×10$^6$ | ~1.4×10$^5$ |
| Composite cell 2 | ~40×35×35 | 7 | 14.0 | ~3.5×10$^6$ | ~3.4×10$^5$ |

**TABLE S2.** Calculated bulk and defect properties of pure Nb using the present 2NN MEAM potential, in comparison with experimental data, DFT data, and previous MEAM calculations by Lee *et al.* [6]. Following quantities are listed: the cohesive energy $E_c$ (eV/atom), the lattice constant $a$ (Å), the bulk modulus $B$ and the elastic constants $C_{11}$, $C_{12}$ and $C_{44}$ (10$^{12}$ dyne/cm$^2$), structural energy differences $\Delta E$ (eV/atom), the vacancy formation energy $E_f^{vac}$ (eV), the vacancy migration energy $E_m^{vac}$ (eV), the activation energy of vacancy diffusion $Q^{vac}$ (eV), and the surface energies $E_{surf}$ (erg/cm$^2$) for the orientations indicated by the superscript.

| Property | Exp. [a] | DFT [b] | 2NN MEAM [Lee [a]] | 2NN MEAM [This work] |
|---|---|---|---|---|
| $E_c$ | 7.47 | 6.957 | 7.47 | 7.47 |
| $a$ | 3.303 | 3.323 | 3.302 | 3.302 |
| $B$ | 1.73 | 1.722 | 1.73 | 1.762 |
| $C_{11}$ | 2.527 | 2.460 | 2.527 | 2.567 |
| $C_{12}$ | 1.332 | 1.365 | 1.331 | 1.360 |
| $C_{44}$ | 0.310 | 0.298 | 0.319 | 0.303 |
| $\Delta E_{bcc \to fcc}$ | | 0.320 | 0.176 | 0.189 |
| $\Delta E_{bcc \to hcp}$ | | 0.295 | 0.164 | 0.214 |
| $E_f^{vac}$ | 2.75 | 2.518 | 2.75 | 3.202 |
| $E_m^{vac}$ | | 0.414 | 0.57 | 0.425 |
| $Q^{vac}$ | 3.6 | 2.931 | 3.32 | 3.627 |
| $E_{surf}^{(100)}$ | 2300 [c], 2983 [c] | 2271 | 2715 | 2478 |
| $E_{surf}^{(110)}$ | | 2062 | 2490 | 2092 |
| $E_{surf}^{(111)}$ | | 2322 | 2923 | 2641 |

[a] Ref. [6] and references therein.
[b] Present DFT calculations.
[c] The experimental value is for a polycrystalline solid.

**TABLE S3.** Calculated thermal properties of the pure Nb system using the present 2NN MEAM potential, in comparison with experimental data and previous calculations by Lee *et al.* [6]. The listed quantities correspond to the thermal expansion coefficient $\varepsilon$ ($10^{-6}$/K), the heat capacity at constant pressure $C_P$ (J/mol K), the melting temperature $T_m$ (K), the enthalpy of melting $\Delta H_m$ (kJ/mol), and the volume change upon melting $\Delta V_m/V_{\text{solid}}$ (%).

| System | Property | Experiment [a] | 2NN MEAM [Lee [a]] | 2NN MEAM [This work] |
|---|---|---|---|---|
| Nb (bcc) | $\varepsilon$ (0–100 °C) | 7.2 | 6.4 | 5.6 |
| | $C_P$ (0–100 °C) | 24.9 | 26.1 | 25.5 |
| | $T_m$ | 2750 | 1900 | 2665 |
| | $\Delta H_m$ | 30.0 | 13.5 | 21.0 |
| | $\Delta V_m/V_{\text{solid}}$ | | 1.0 | 2.8 |

[a] Ref. [6] and references therein.

**TABLE S4**. Calculated physical properties (in units of eV) of the binary solid solutions using the present 2NN MEAM potential, in comparison with DFT. Following quantities are listed: the dilute heat of solution $E^{sol}$, the vacancy-solute binding energy $E_b^{vac-sol}$, the solute-solute binding energy $E_b^{sol-sol}$, and the solute migration energy $E_m^{sol}$. The reference states for the dilute heat of solution are bcc Nb, fcc Ni, and hcp Ti. In the bcc and fcc phases, the energies for first and second nearest neighbor bindings are designated by "1NN" and "2NN", respectively. In the hcp phase, in-basal plane and out-of-basal plane binding and migration energies are designated by "in" and "out", respectively. The defect binding energies are defined such that a positive sign indicates attractive interaction.

| System | Structure | Property | DFT [a] | 2NN MEAM |
|---|---|---|---|---|
| Nb-Ni | Nb-rich bcc | $E^{Ni}$ | -0.076 | -0.148 |
| | | $E_b^{vac-Ni}$(1NN) | 0.067 | 0.063 |
| | | $E_b^{vac-Ni}$(2NN) | 0.338 | 0.332 |
| | | $E_b^{Ni-Ni}$(1NN) | -0.048 | -0.078 |
| | | $E_b^{Ni-Ni}$(2NN) | 0.160 | 0.147 |
| | | $E_m^{Ni}$ | 0.982 | 1.051 |
| | Ni-rich fcc | $E^{Nb}$ | -0.592 | -0.127 |
| | | $E_b^{vac-Nb}$(1NN) | 0.127 | 0.120 |
| | | $E_b^{vac-Nb}$(2NN) | -0.064 | 0.048 |
| | | $E_b^{Nb-Nb}$(1NN) | -0.487 | -0.243 |
| | | $E_b^{Nb-Nb}$(2NN) | 0.207 | 0.249 |
| | | $E_m^{Ti}$ | 0.702 | 0.890 |
| Nb-Ti | Nb-rich bcc | $E^{Ti}$ | 0.062 | -0.095 |
| | | $E_b^{vac-Ti}$(1NN) | 0.157 | 0.305 |
| | | $E_b^{vac-Ti}$(2NN) | 0.095 | -0.079 |
| | | $E_b^{Ti-Ti}$(1NN) | 0.016 | 0.028 |
| | | $E_b^{Ti-Ti}$(2NN) | -0.027 | -0.037 |
| | | $E_m^{Ti}$ | 0.375 | 0.511 |
| | Ti-rich hcp | $E^{Nb}$ | 0.336 | 0.529 |
| | | $E_b^{vac-Nb}$(in) | 0.098 | 0.129 |
| | | $E_b^{vac-Nb}$(out) | 0.079 | 0.042 |
| | | $E_b^{Nb-Nb}$(in) | -0.004 | 0.109 |
| | | $E_b^{Nb-Nb}$(out) | -0.030 | 0.076 |
| | | $E_m^{Nb}$(in) | 0.216 | 0.571 |
| | | $E_m^{Nb}$(out) | 0.207 | 0.409 |

[a] Present DFT calculation.

**TABLE S5**. Calculated physical properties of binary compounds at various compositions using the present 2NN MEAM potential, in comparison with experimental and DFT data. Following quantities are listed: the lattice constants $a$, $b$, and $c$ (Å), the bulk modulus $B$ ($10^{12}$ dyne/cm$^2$), and the enthalpy of formation $\Delta E_f$ (eV/atom).

| System | Composition | Structure | Stability | Property | Exp.[a] | DFT[b] | 2NN MEAM |
|---|---|---|---|---|---|---|---|
| Nb-Ni | NbNi$_3$ | D0$_a$ | Stable | $a$ | 5.116 | 5.125 | 5.217 |
| | | | | $b$ | 4.565 | 4.566 | 4.810 |
| | | | | $c$ | 4.258 | 4.259 | 4.199 |
| | | | | $B$ | | 2.073 | 2.152 |
| | | | | $\Delta E_f$ | | -0.295 | -0.251 |
| | | L1$_2$ | Hypothetical | $a$ | | 3.694 | 3.755 |
| | | | | $B$ | | 1.993 | 2.100 |
| | | | | $\Delta E_f$ | | -0.134 | -0.208 |
| | NbNi$_2$ | C1 | Hypothetical | $a$ | | 5.768 | 6.016 |
| | | | | $B$ | | 1.423 | 1.274 |
| | | | | $\Delta E_f$ | | 0.431 | 0.732 |
| | NbNi | B1 | Hypothetical | $a$ | | 5.042 | 5.222 |
| | | | | $B$ | | 1.666 | 1.403 |
| | | | | $\Delta E_f$ | | 0.372 | 1.040 |
| | | B2 | Hypothetical | $a$ | | 3.088 | 3.120 |
| | | | | $B$ | | 1.917 | 2.137 |
| | | | | $\Delta E_f$ | | -0.018 | -0.276 |
| | Nb$_7$Ni$_6$ | D8$_5$ | Stable | $a$ | 4.893 | 4.942 | 4.962 |
| | | | | $c$ | 26.64 | 27.073 | 27.219 |
| | | | | $B$ | | 1.975 | 2.119 |
| | | | | $\Delta E_f$ | | -0.207 | -0.295 |
| | Nb$_2$Ni | C1 | Hypothetical | $a$ | | 5.968 | 5.993 |
| | | | | $B$ | | 1.561 | 1.426 |
| | | | | $\Delta E_f$ | | 0.345 | 0.632 |
| | Nb$_3$Ni | L1$_2$ | Hypothetical | $a$ | | 4.052 | 4.013 |
| | | | | $B$ | | 1.738 | 1.478 |
| | | | | $\Delta E_f$ | | 0.103 | 0.021 |

| | | | | | | |
|---|---|---|---|---|---|---|
| Nb-Ti | NbTi₃ | L1₂ | Hypothetical | $a$ | 4.132 | 4.138 |
| | | | | $B$ | 1.233 | 1.011 |
| | | | | $\Delta E_f$ | 0.080 | 0.048 |
| | NbTi₂ | C1 | Hypothetical | $a$ | 6.229 | 6.078 |
| | | | | $B$ | 0.923 | 0.675 |
| | | | | $\Delta E_f$ | 0.747 | 0.768 |
| | NbTi | B1 | Hypothetical | $a$ | 5.345 | 5.404 |
| | | | | $B$ | 1.073 | 0.571 |
| | | | | $\Delta E_f$ | 0.851 | 1.460 |
| | | B2 | Hypothetical | $a$ | 3.271 | 3.270 |
| | | | | $B$ | 1.369 | 1.049 |
| | | | | $\Delta E_f$ | 0.070 | 0.018 |
| | Nb₂Ti | C1 | Hypothetical | $a$ | 6.228 | 6.442 |
| | | | | $B$ | 1.133 | 0.612 |
| | | | | $\Delta E_f$ | 0.883 | 1.486 |
| | Nb₃Ti | L1₂ | Hypothetical | $a$ | 4.196 | 4.161 |
| | | | | $B$ | 1.542 | 1.250 |
| | | | | $\Delta E_f$ | 0.217 | 0.115 |

[a] Ref. [7].
[b] Present DFT calculations.

**TABLE S6**. Calculated physical properties (in units of eV) of the ternary solid solutions using the present 2NN MEAM potential, in comparison with DFT. Following quantities are listed: solute-solute binding energy $E_b^{sol-sol}$. In the bcc and fcc phases, the energies for first and second nearest neighbor bindings are designated by "1NN" and "2NN", respectively. In the hcp phase, in-basal plane and out-of-basal plane binding and migration energies are designated by "in" and "out", respectively. The defect binding energies are defined such that a positive sign indicates attractive interaction.

| System | Structure | Property | DFT [a] | 2NN MEAM [This work] |
|---|---|---|---|---|
| Nb-Ni-Ti | bcc Nb | $E_b^{Ni-Ti}$(1NN) | 0.119 | 0.096 |
| | | $E_b^{Ni-Ti}$(2NN) | -0.008 | -0.040 |
| | fcc Ni | $E_b^{Nb-Ti}$(1NN) | -0.433 | -0.403 |
| | | $E_b^{Nb-Ti}$(2NN) | 0.147 | 0.016 |
| | hcp Ti | $E_b^{Nb-Ni}$(in) | 0.079 | 0.015 |
| | | $E_b^{Nb-Ni}$(out) | 0.050 | 0.002 |

[a] Present DFT calculation.

**TABLE S7**. Calculated physical properties of hypothetical Cu$_2$MnAl-type ternary compounds using the present 2NN MEAM potential, in comparison with experimental and DFT data. Following quantities are listed: the lattice constants *a, b,* and *c* (Å), the bulk modulus *B* ($10^{12}$ dyne/cm$^2$), and the enthalpy of formation $\Delta E_f$ (eV/atom).

| System | Structure | Property | DFT [a] | 2NN MEAM [This work] |
|---|---|---|---|---|
| Nb-Ni-Ti | Nb$_2$NiTi | *a* | 6.333 | 6.362 |
| | | *B* | 1.616 | 1.472 |
| | | $\Delta E_f$ | 0.050 | -0.026 |
| | NbNi$_2$Ti | *a* | 6.103 | 6.155 |
| | | *B* | 1.757 | 1.665 |
| | | $\Delta E_f$ | -0.181 | -0.336 |
| | NbNiTi$_2$ | *a* | 6.256 | 6.171 |
| | | *B* | 1.423 | 1.268 |
| | | $\Delta E_f$ | -0.114 | -0.275 |

[a] Present DFT calculation.

# S-3. Supplementary figures

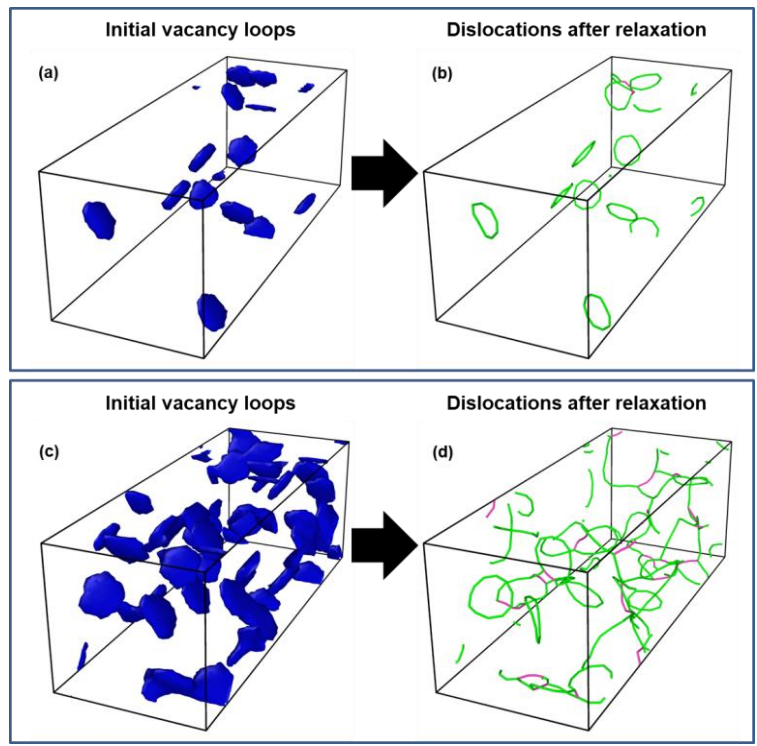

**Fig. S1.** Evolution of the dislocation structure inside of the Nb nanowire; note that a cuboidal box with dimensions of ~20×7.5×7.3 nm is shown, in which the Nb nanowire cylinder (diameter of 7 nm) is contained. The diameters of the vacancy loops are 3.5 for (a, b) and 4 for (c, d) times the lattice parameter (0.33 nm). The atomic configurations before (a, c) and after (b, d) the relaxation are shown. In the dislocation structures after the relaxation, the green lines correspond to $\frac{1}{2}$<111> and magenta lines correspond to <100> dislocations.

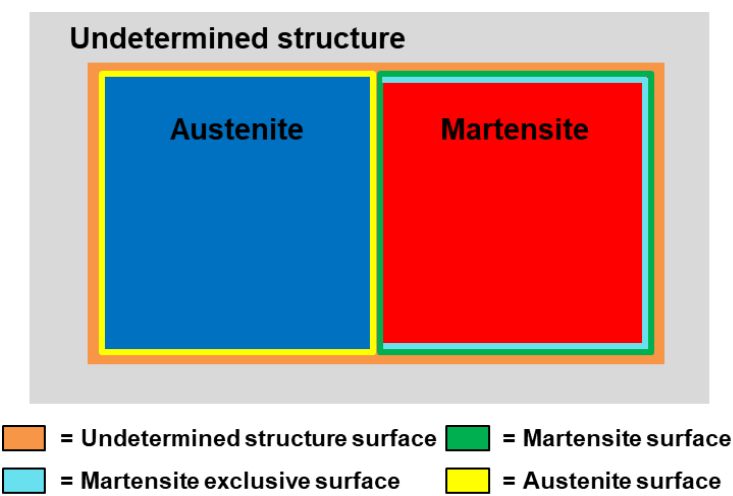

**Fig. S2.** Schematic diagram illustrating the austenite, martensite, undetermined structure, and martensite exclusive surfaces with designated layers. The legend at the bottom explains the color coding for the surface layers.

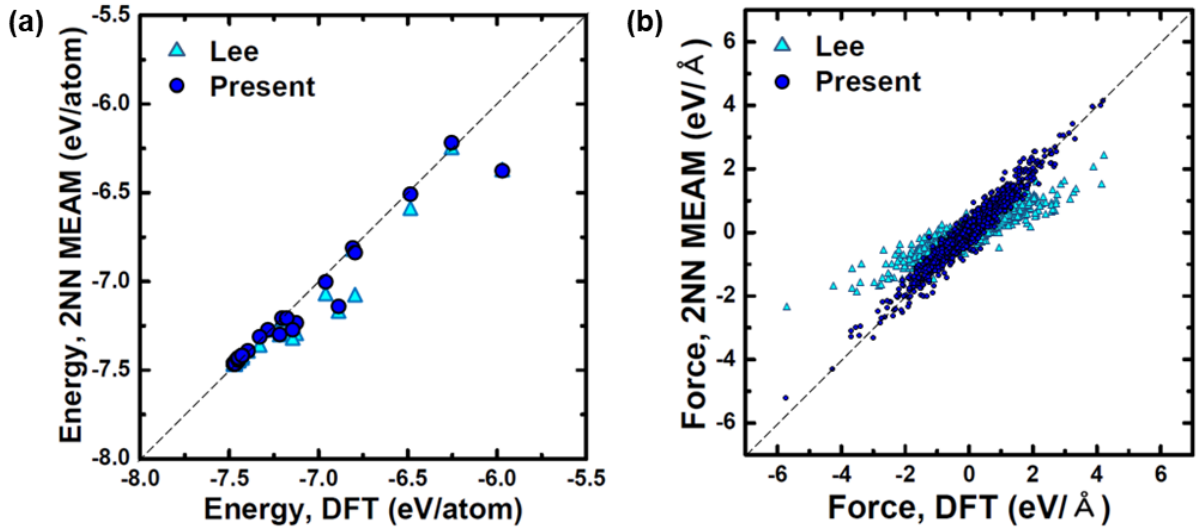

**Fig. S3.** Scatter plots for (a) energies and (b) forces of pure Nb with respect to the DFT database. Values obtained with the present potential are compared to those obtained with the previous potential by Lee *et al.* (Ref. [6]). A perfect correlation with the DFT values would correspond to the dashed lines.

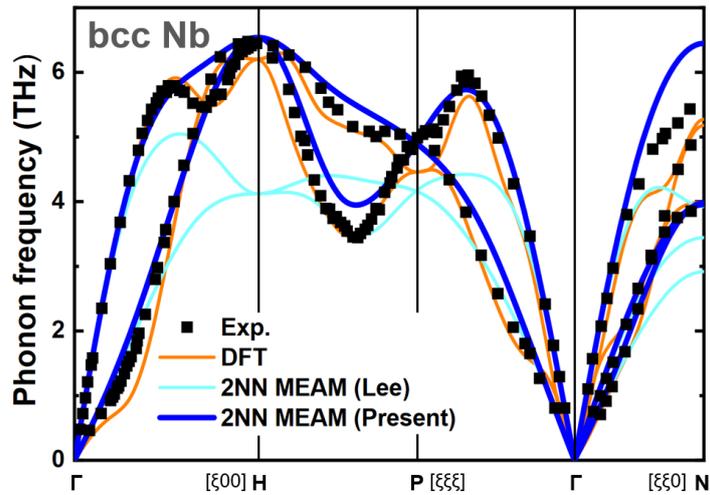

**Fig. S4.** Calculated phonon spectra of the bcc Nb phase using the present 2NN MEAM potential, in comparison with experimental data (Ref. [8]), DFT results, and previous calculations by Lee *et al.* (Ref. [6]).

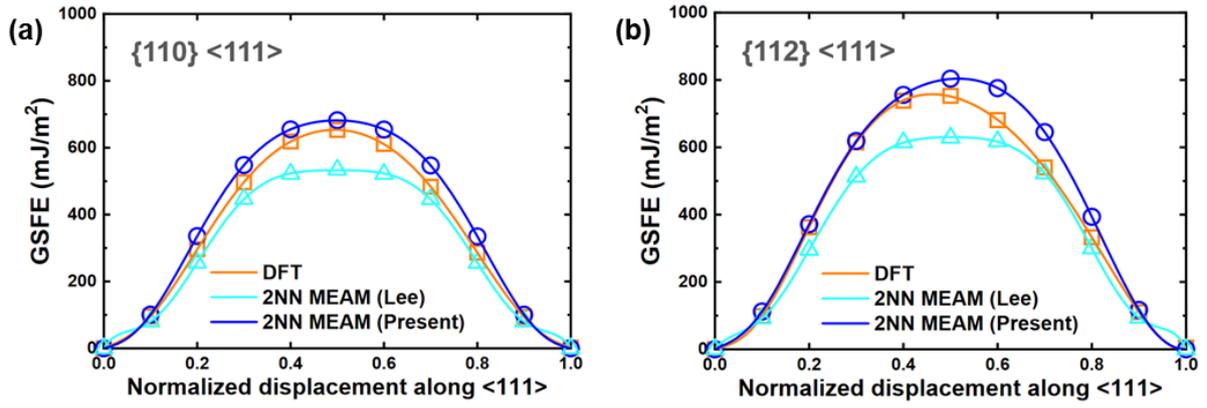

**Fig. S5.** Calculated GSFE curves of bcc Nb for <111> dislocations on the (a) {110} and (b) {112} slip planes using the present 2NN MEAM potential, in comparison with present DFT results and previous calculations by Lee et al. (Ref. [6]).

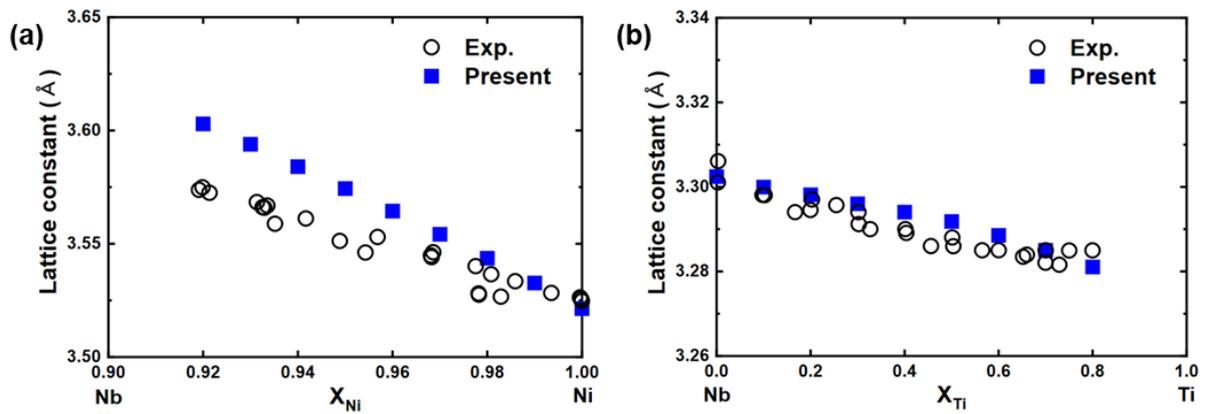

**Fig. S6.** Composition dependence of the lattice constant of the (a) Nb-Ni (Ni-rich) fcc and (b) Nb-Ti (Nb-rich) bcc solid solutions calculated using the present 2NN MEAM potential, in comparison with experimental data (Refs. [7, 9]).

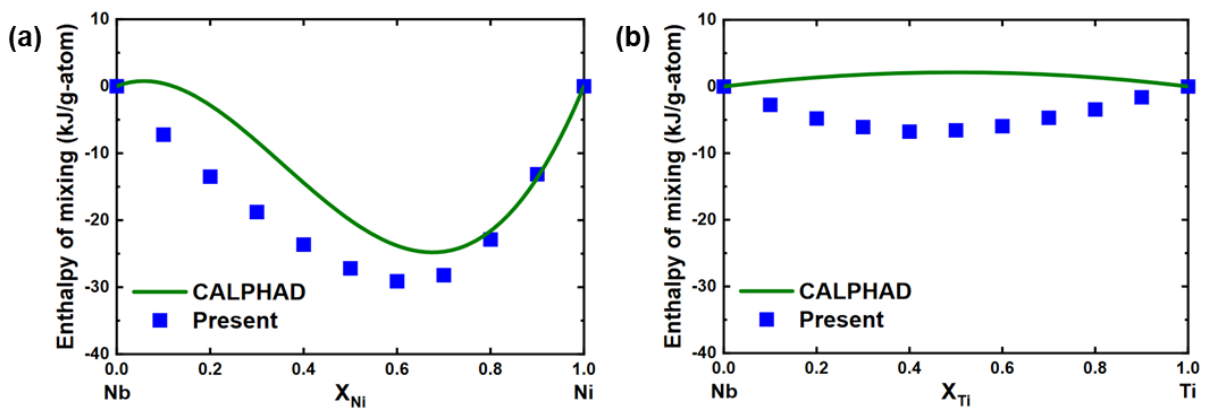

**Fig. S7.** Enthalpy of mixing of the (a) Nb-Ni and (b) Nb-Ti liquid phase at 2800 K calculated using the present 2NN MEAM potential, in comparison with CALPHAD based modeling (Ref. [10]).

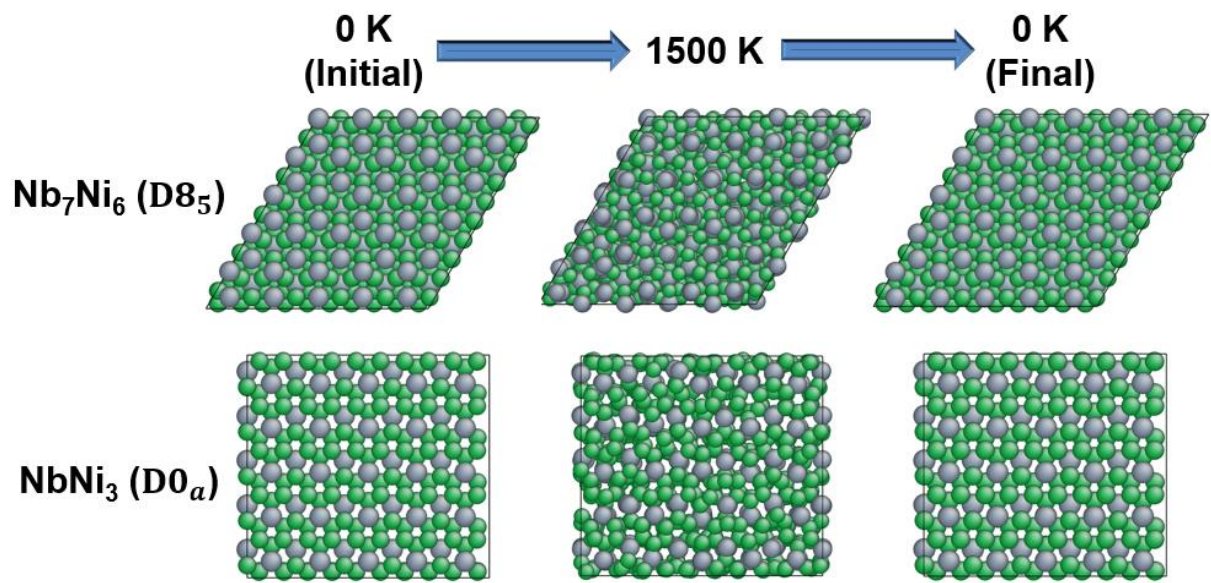

**Fig. S8.** Stability of intermetallic compounds (D8$_5$-Nb$_7$Ni$_6$ and D0$_a$-NbNi$_3$) examined by MD simulations based on the present 2NN MEAM potential. Atomic snapshots obtained during the thermal loading process (0 K → 1500 K → 0 K) are presented. Ni and Nb atoms are represented by green and gray balls, respectively.

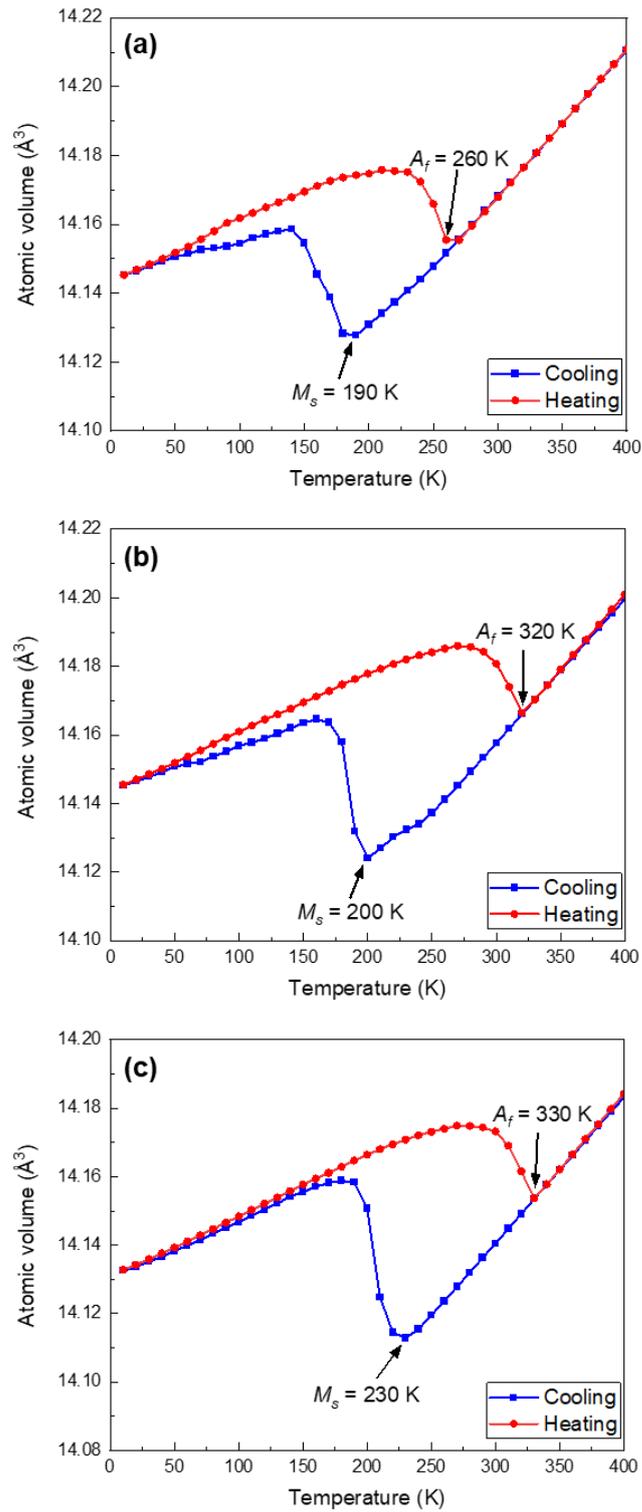

**Fig. S9.** Temperature dependence of the atomic volume of the composite cells with dimensions of (a) 20×17.5×17.5 nm, (b) 30×26×26 nm, and (c) 40×35×35 nm during cooling (400 → 10 K) and reheating (10 → 400 K). The martensite start ($M_s$) and austenite finish ($A_f$) temperatures are indicated by the arrows.

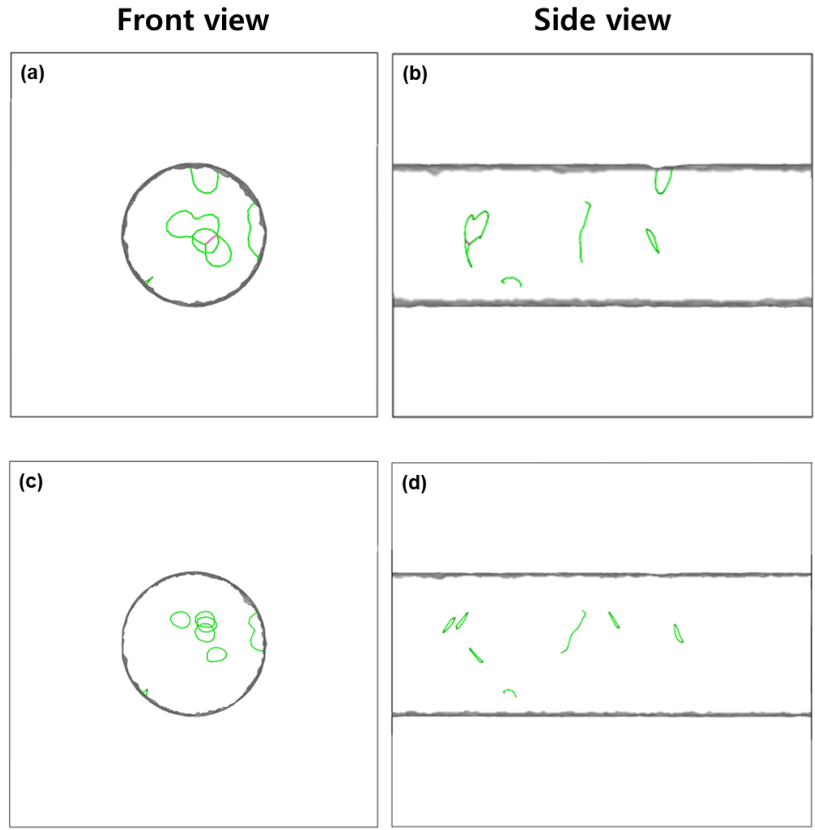

**Fig. S10.** Front and side views of composite cells showing dislocations after the relaxation at 300 K. (a-b) cell with dimensions of 30×26×26 nm and dislocation density of 0.014 nm$^{-2}$. (c-d) cell with dimensions of 40×35×35 nm and dislocation density of 0.006 nm$^{-2}$. Green lines correspond to $\frac{1}{2}$<111> and magenta lines correspond to <100> dislocations.

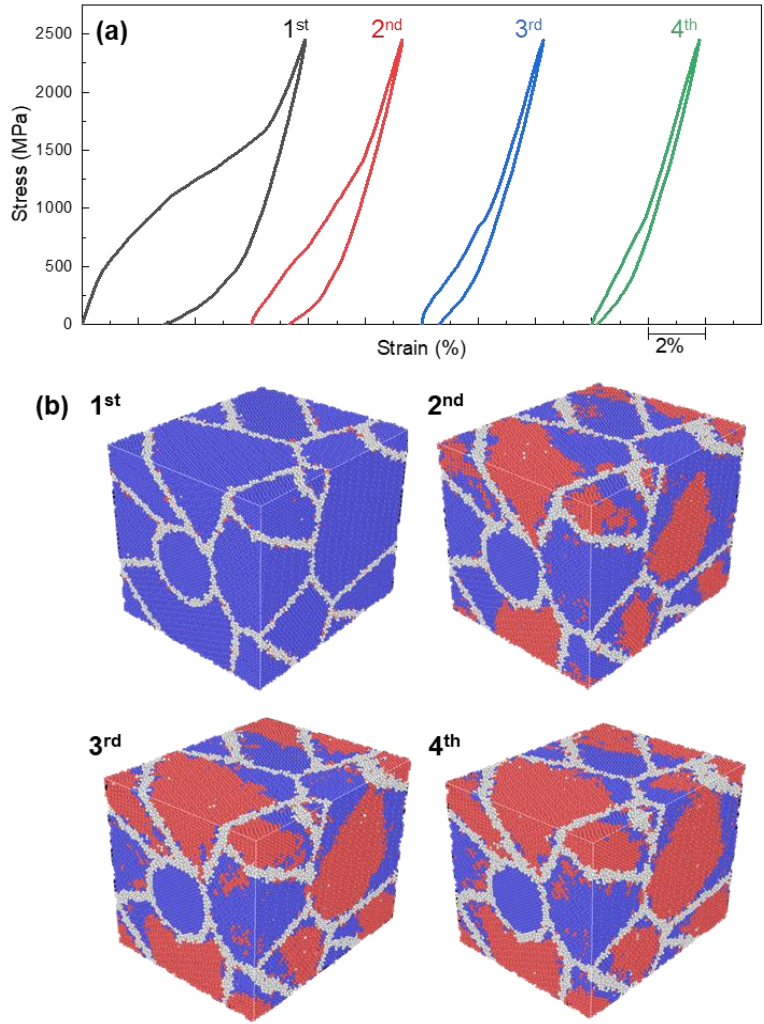

**Fig. S11.** (a) Cyclic stress-strain curves of the Nb nanowire + NiTi SMA composite with cell dimensions of 20×17.5×17.5 nm and initial dislocation density of 0.04 nm$^{-2}$, obtained by an MD simulation at 300 K with a stress rate of 8.33 MPa/ps. (b) Atomic configurations at the initial stage of each cyclic loading. The blue atoms correspond to the austenite (B2) structure, red atoms to the martensite (B19') structure, and gray atoms represent undetermined structures (grain boundary, wire-matrix interface, amorphous region, and thermal noise).

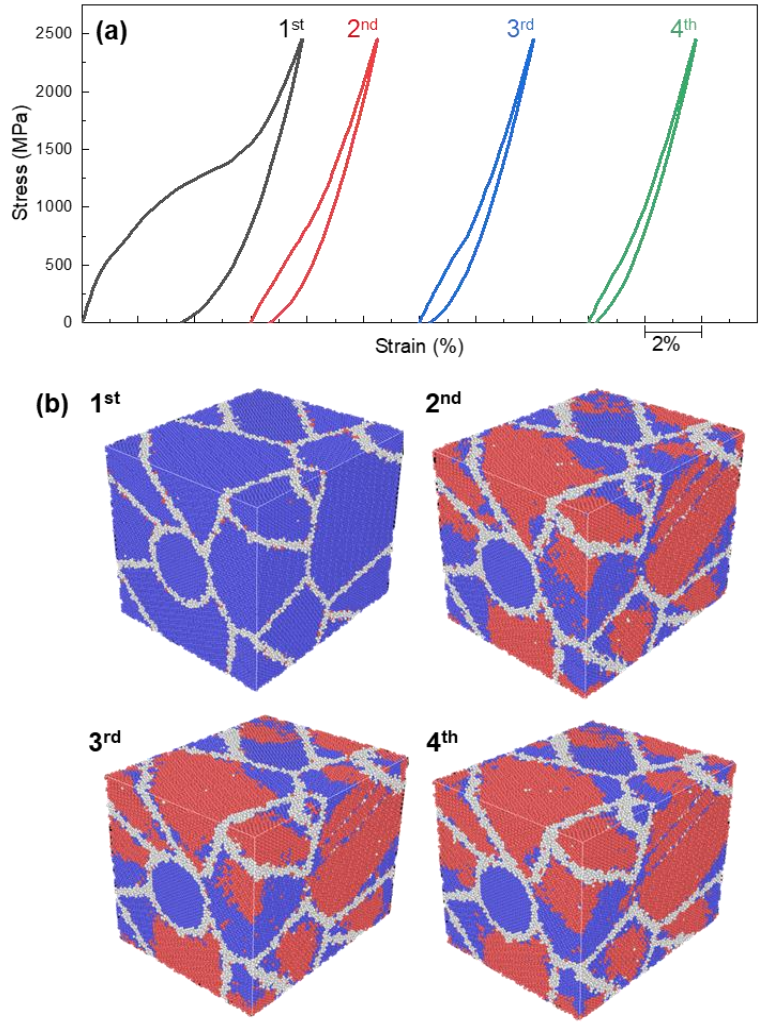

**Fig. S12.** (a) Cyclic stress-strain curves of the Nb nanowire + NiTi SMA composite with cell dimensions of 20×17.5×17.5 nm and initial dislocation density of 0.04 nm$^{-2}$, obtained by an MD simulation at 300 K with a stress rate of 12.5 MPa/ps. (b) Atomic configurations at the initial stage of each cyclic loading. The blue atoms correspond to the austenite (B2) structure, red atoms to the martensite (B19') structure, and gray atoms represent undetermined structures (grain boundary, wire-matrix interface, amorphous region, and thermal noise).

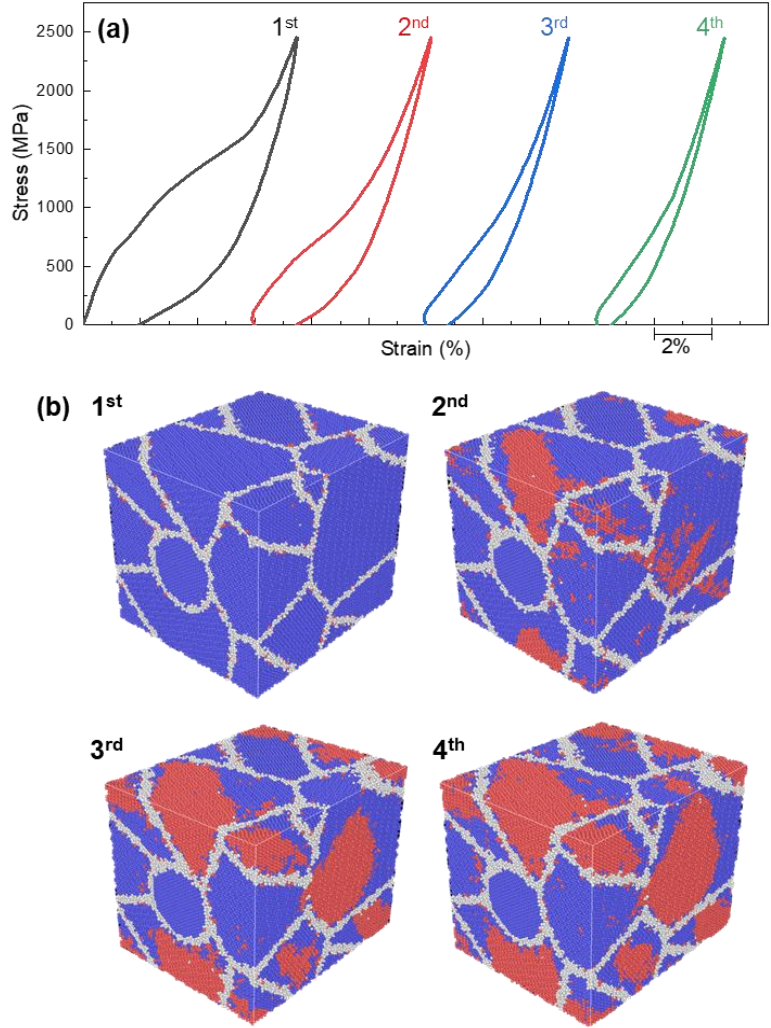

**Fig. S13.** (a) Cyclic stress-strain curves of the Nb nanowire + NiTi SMA composite with cell dimensions of 20×17.5×17.5 nm and initial dislocation density of 0.04 nm$^{-2}$, obtained by an MD simulation at 300 K with a stress rate of 25.0 MPa/ps. (b) Atomic configurations at the initial stage of each cyclic loading. The blue atoms correspond to the austenite (B2) structure, red atoms to the martensite (B19') structure, and gray atoms represent undetermined structures (grain boundary, wire-matrix interface, amorphous region, and thermal noise).

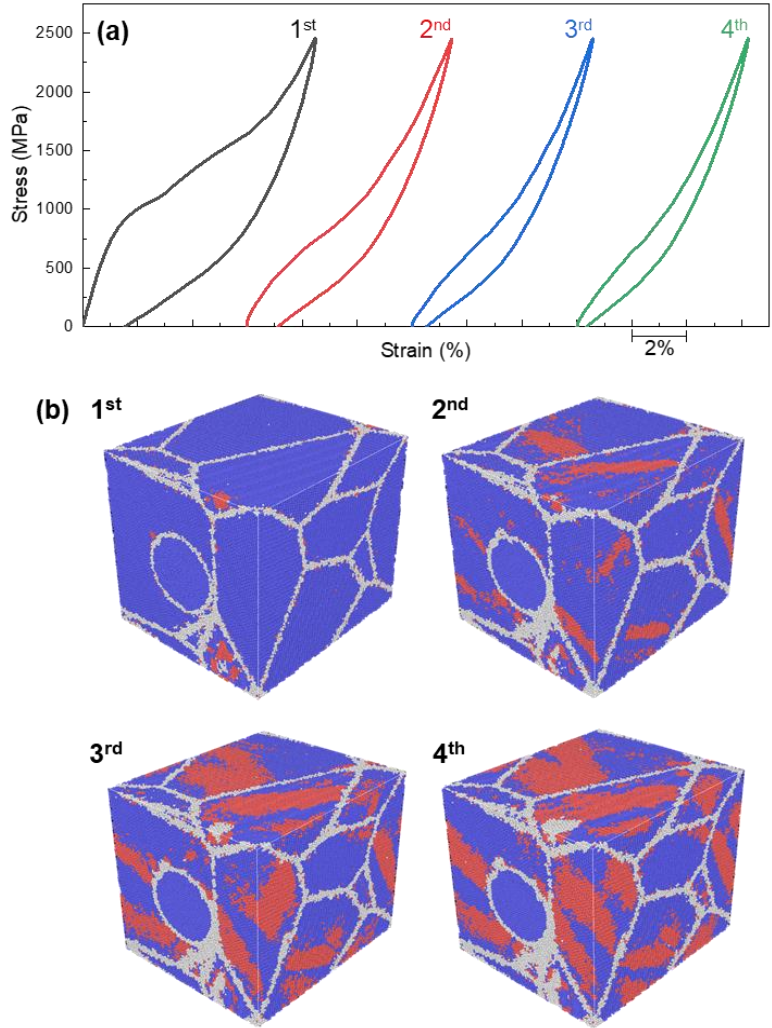

**Fig. S14.** (a) Cyclic stress-strain curves of the Nb nanowire + NiTi SMA composite with cell dimensions of 30×26×26 nm and initial dislocation density of 0.014 nm$^{-2}$, obtained by an MD simulation at 340 K with a stress rate of 12.5 MPa/ps. (b) Atomic configurations at the initial stage of each cyclic loading. The blue atoms correspond to the austenite (B2) structure, red atoms to the martensite (B19') structure, and gray atoms represent undetermined structures (grain boundary, wire-matrix interface, amorphous region, and thermal noise).

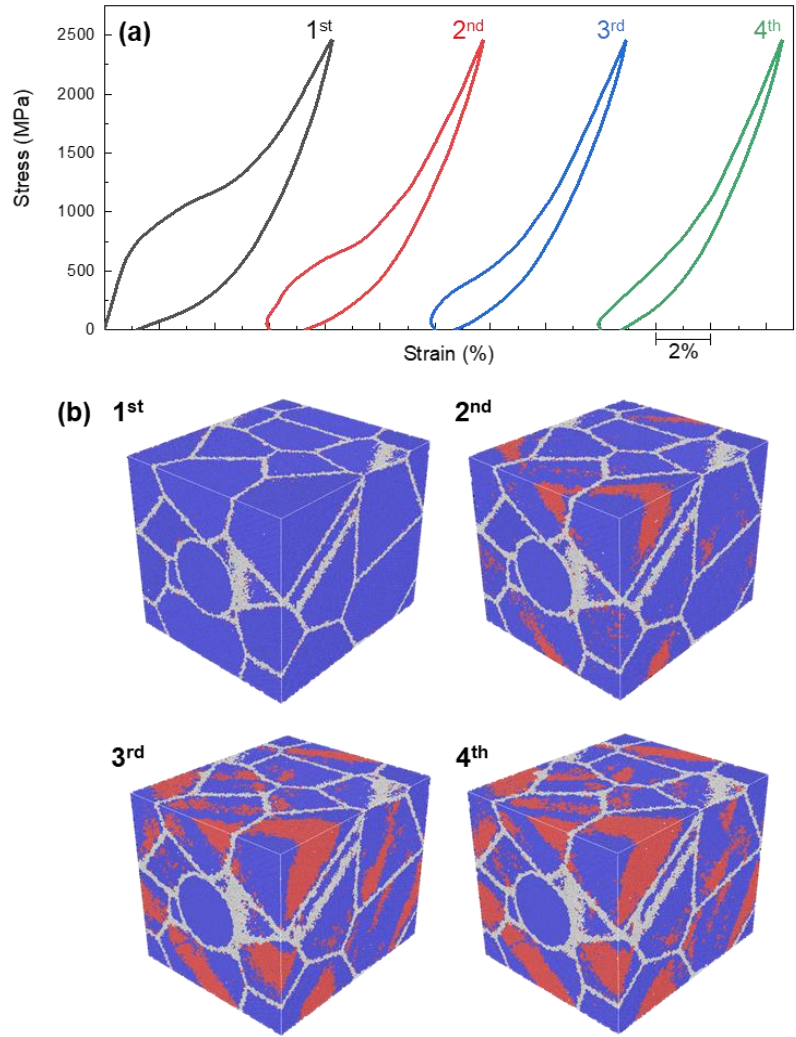

**Fig. S15.** (a) Cyclic stress-strain curves of the Nb nanowire + NiTi SMA composite with cell dimensions of 40×35×53 nm and initial dislocation density of 0.006 nm$^{-2}$, obtained by an MD simulation at 350 K with a stress rate of 12.5 MPa/ps. (b) Atomic configurations at the initial stage of each cyclic loading. The blue atoms correspond to the austenite (B2) structure, red atoms to the martensite (B19') structure, and gray atoms represent undetermined structures (grain boundary, wire-matrix interface, amorphous region, and thermal noise).

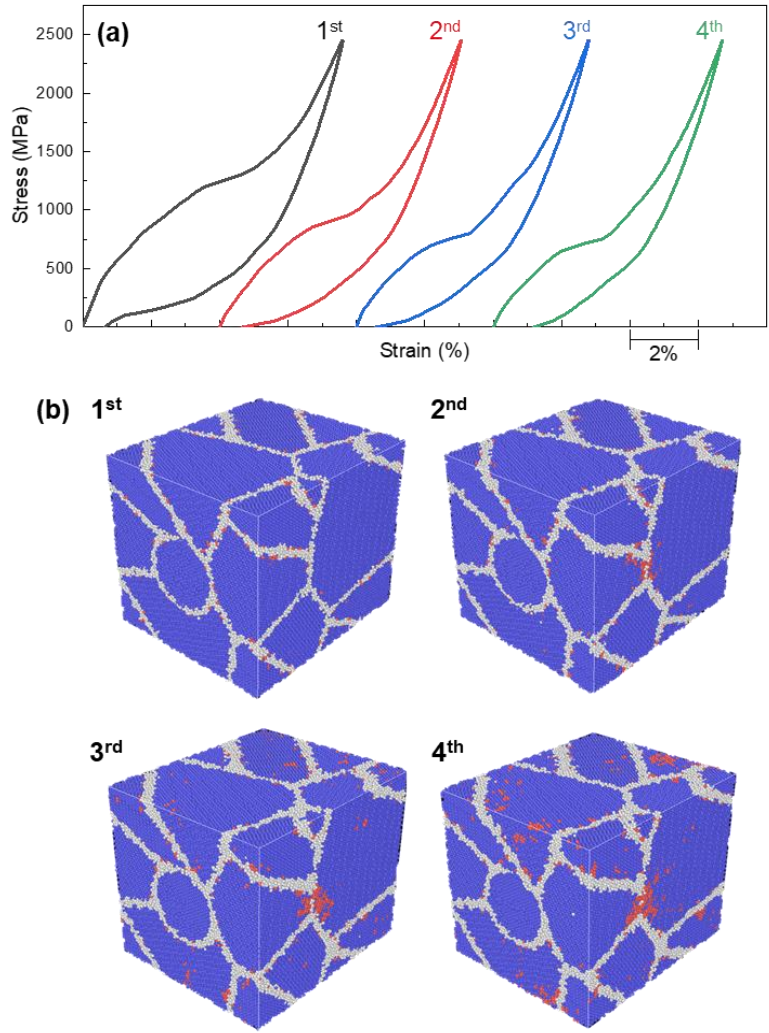

**Fig. S16.** (a) Cyclic stress-strain curves of the Nb nanowire + NiTi SMA composite with cell dimensions of 20×17.5×17.5 nm without dislocations, obtained by an MD simulation at 300 K with a stress rate of 6.25 MPa/ps. (b) Atomic configurations at the initial stage of each cyclic loading. The blue atoms correspond to the austenite (B2) structure, red atoms to the martensite (B19') structure, and gray atoms represent undetermined structures (grain boundary, wire-matrix interface, amorphous region, and thermal noise).